\documentclass[11pt,a4paper]{article}
\usepackage{format}

\title{A Minimum Distance Estimator Approach for Misspecified Ergodic Processes\thanks{Submitted to \emph{SIAM Multiscale Modeling and Simulation (MMS)} on June 14, 2025.}}
\author{Jaroslav I. Borodavka, Sebastian Krumscheid and Grigorios A. Pavliotis}
\date{\today}

\begin{document}
\maketitle

\begin{abstract}
\noindent
We propose a minimum distance estimator (MDE) for parameter identification in misspecified models characterized by a sequence of ergodic stochastic processes that converge weakly to the model of interest. The data is generated by the sequence of processes, and we are interested in inferring parameters for the limiting processes. We define a general statistical setting for parameter estimation under such model misspecification and prove the robustness of the MDE. Furthermore, we prove the asymptotic normality of the MDE for multiscale diffusion processes with a well-defined homogenized limit. A tractable numerical implementation of the MDE is provided and realized in the programming language Julia.
\end{abstract}

\section{Introduction}
In the classical statistics literature, the topic of parameter estimation is one of the main building blocks of statistical inference for stochastic processes. It is a reoccurring theme to study such inference tasks in a specified model setting, that is, when the data is assumed to come from one particular parametric family of distributions, see \cite{IK:1981, K:2004, LS:2000, R:1999}. This setting undoubtedly holds up relevant, but is often violated in applications where a mismatch between data and model is common. One prominent example of such a mismatch can be found in the realm of multiscale modeling where complicated dynamical systems, characterized through processes evolving on multiple timescales, are approximated by a surrogate model capturing effective dynamics of the full system; refer to the references given in the second paragraph. When using a method such as scale separation, then this is commonly known as coarse-graining, and multiscale analysis is termed averaging or homogenization; see \cite{BLP:1978, FW:2012, PS:2008, RX:2021}. Yet, even after having derived a surrogate model, it is often times not possible to give closed-form expressions for the relevant model parameters analytically. One promising way to still obtain a surrogate model in such a situation is to resort to data, which, however, is compatible with the surrogate model only at the appropriate macroscopic length and time scales. This is a particular setting of misspecification between data and model, cf. \cite[Sec. 2.6.1]{K:2004}, and is practically relevant for two main reasons. One reason is the imperfect knowledge of the data-generating model. After all, a surrogate model may very well approximate several different data-generating models, which are unknown, in general. The second reason is that, even if there is perfect knowledge of the data-generating model, then the computational complexity for simulation or estimation is so high that it becomes inconvenient or just impossible to work with the data-generating model. Hence, from a practitioner's point of view, it may be desirable to work with a simpler model. \\ \indent 
This multiscale modeling procedure is an established practice in various applications, for example, in econometrics when dealing with high-frequency financial data and market microstructure \cite{T:2005}, in network traffic data \cite{ABFRV:2002}, in molecular dynamics to find reaction coordinates and collective variables \cite{ABJT:2013, S:2010}, or in the study of intracellular biochemical reaction networks involving multiple reactions and chemical species \cite{BKPR:2006, KRK:2017}. We refer interested readers to the books \cite{K:2015, PS:2008} for a more complete overview. \\ \indent
In this work, we explore parameter estimation for misspecified
models where a sequence of ergodic stochastic processes, assumed to come from the data-generating model and characterized by a small-scale parameter, converges weakly to a stochastic process that describes the surrogate model. This kind of "weak" mismatch between the data-generating model and the surrogate model appears to be difficult to solve robustly with standard statistical estimators. For example, in \cite{PPS:2009, PS:2007} the authors rigorously prove that the maximum likelihood estimator, if built on the basis of the coarse-grained model, cannot recover the true parameter when it is fed with multiscale observations coming from the slow-evolving component of the full multiscale system. This is peculiar in the sense that, when the small-scale parameter goes to zero, the underlying slow-evolving process of the multiscale model converges weakly to the process describing the coarse-grained model. Intuition tells us that such a "weak" misspecification should not lead to a significant bias in the estimation. \\ \indent 
In recent decades, much effort has been put into answering and solving this counterintuitive issue. For example, in \cite{PPS:2009, PS:2007} a subsampling approach was proposed in which an appropriate parameter-dependent subsampling rate is used to infer the true parameters through the maximum likelihood estimator. Subsampling, together with averaging, has also been applied in the context of market microstructure and high-frequency financial data, for example, \cite{AMZ:2005, ZMA:2003}. Another approach was introduced in \cite{KPK:2013, KKP:2015, K:2018} in a semi-parametric setting that uses the martingale properties of diffusion processes as estimating equations. Recently, a very promising approach based on filtered data has been consistently applied to several settings including multiscale Langevin dynamics \cite{AGPSZ:2021}, discretely observed multiscale diffusions \cite{APZ:2022}, and stochastic processes driven by colored noise \cite{PRZ:2025}. In the last mentioned article, the authors also considered, in addition to the filtering approach, a stochastic gradient descent in continuous time as an estimator. \\ \indent 
In this work, we want to tackle the problem motivated by the following heuristic idea. Given certain assumptions on the underlying processes, most and foremost weak convergence, it should be possible to retrieve the true parameter of the limiting model robustly from multiscale observations through quantities that converge in a suitable metric. This idea lends itself to a minimum distance approach, which we want to employ and present under various theoretical and numerical aspects. In particular, our approach is inspired by the classical BHEP goodness-of-fit test for multivariate normality, which was thoroughly studied in \cite{HW:1997} and later generalized to the setting of random elements in a separable Hilbert space in \cite{HJ:2021}. Minimum distance estimation is an ubiquitous method for statistical inference in parametric problems that, in certain contexts, encapsulates many commonly known inference methods as special cases, e.g., maximum likelihood estimation, least squares, and method of moments. The methodology was originally developed and introduced by Wolfowitz in \cite{W:1953}. Since then, the method has been extensively studied and applied; see, for example, \cite{K:1994} and the references therein for a comprehensive analysis of minimum distance estimators (MDEs) in the context of statistical inference for diffusion processes. Although MDEs enjoy several favorable theoretical properties such as consistency or asymptotic normality, a prevailing shortcoming is their computational cost; cf. \cite{CM:2011} for a brief discussion of this aspect. We will partially address this issue in the following by providing a computationally tractable formula for the MDE in a multivariate Gaussian case and, in non-Gaussian settings, by utilizing a fast Fourier transform algorithm for the efficient and accurate approximation of appearing convolution integrals. \\ \indent
Our main contributions can be summarized as follows. We provide an alternative estimator for the problem of parameter identification for misspecified models that have a certain "weak" misspecification between the data and the model. The MDE used can be computed efficiently in a multivariate Gaussian case. We define a statistical estimation framework for misspecified models and prove several asymptotic properties of the MDE, e.g., robustness under the true parameter, for different limiting regimes under general assumptions. We also provide a rigorous proof for the asymptotic normality of the MDE in a misspecified setting where the time horizon explicitly depends on the small-scale parameter of the multiscale system.

The remainder of the paper is organized as follows. We describe the setting and propose the approach in Section \ref{sec:setting}. In Section \ref{sec:asymptotic_properties} we introduce a general estimation framework for misspecified models, define the notion of robustness, and prove the robustness of the proposed MDE under different limiting regimes. The main results of this section are given in Proposition \ref{prop:robustness_eps_T} and Proposition \ref{prop:robustness_T_eps}. Initial considerations about asymptotic normality are also presented in Proposition \ref{prop:asymptotic_normality}. Then, we analyze the multiscale overdamped Langevin diffusion and its homogenized limit in Section \ref{sec:langevin} under the aspect of robustness and asymptotic normality of the MDE. The main results on asymptotic normality are contained in Proposition \ref{prop:asy_normal_langevin}. Numerical experiments are presented in Section \ref{sec:numerics}, and the theoretical findings are substantiated through a couple of multiscale diffusion examples. Concluding remarks on the features and limitations of the approach, and open questions are given in Section \ref{sec:conclusion}. Finally, Appendix \ref{app:A} contains the required technical results.

\section{Problem Setting and Approach}   \label{sec:setting}
Let $X \colon [0, T] \times \Omega \to \R^d; (t, \omega) \mapsto X(t,\omega)$, $d \in \N$ and $T \in (0, \infty)$, be a continuous-time stochastic process on the probability space $(\Omega, \mathscr{A}, \Pb)$. The first main assumption is that there exists a Lebesgue-density $\mu$ on $\R^d$ such that the convergence
\begin{equation} \label{eq:ergodicity_abstract}
    \lim_{T \rightarrow \infty} \frac1T \int_0^T \phi(X(t)) \, dt = \int_{\R^d} \phi(x) \mu(x) \, dx
\end{equation}
holds $\Pb$-a.s. for $\mu$-integrable functions $\phi \colon \R^d \to \C$, or, in other words, we assume the ergodicity of the process $X$.
For the second main assumption, consider a sequence of stochastic processes $X_\eps \colon [0, T] \times \Omega \to \R^d$ depending on $\eps > 0$. Let the trajectories of $X$ and every $X_\eps$ belong $\Pb$-a.s. to a separable Banach space $\left( \mathscr{X}^T, \| \cdot \|_{\mathscr{X}^T} \right)$ of functions from $[0, T]$ to $\R^d$. We assume that the sequence of push-forward measures $\Pb^{X_\eps}$ converges weakly to $\Pb^{X}$ as $\epstozero$, that is, for all bounded, continuous functions $\Psi \colon \mathscr{X}^T \to \R$ it holds.
\begin{equation} \label{eq:weak_conv}
    \lim_{\epstozero} \int_{\mathscr{X}^T} \Psi(f) \, d \Pb^{X_\eps}(f) = \int_{\mathscr{X}^T} \Psi(f) \, d \Pb^X(f),
\end{equation}
or, equivalently,
\begin{equation*}
    \lim_{\epstozero} \E \Psi(X_\eps) = \E \Psi(X).
\end{equation*}
In this context, we will usually say that $X_\eps$ converges weakly to $X$ in $\mathscr{X}^T$ and call $X_\eps$ a weakly perturbed version of $X$.

Let $\Theta \subseteq \R^p$, $p \in \N$, be an open bounded set. We are interested in estimating a parameter $\vt \in \Theta$ through the given data that manifests itself as a long trajectory of the stochastic process $X$. Assume that this data comes from a parameter-dependent model $\mathcal{M}_\vt$ and that the parameter $\vt$ can be retrieved from the density $\mu$, i.e., $\mu = \mu(\vt) := \mu(\vt, \cdot)$ depends explicitly on the parameter $\vt$. Define the following maps.
\begin{align}
    \label{eq:ec}
    &\ec{T} \colon \mathscr{X}^T \times \R^d \to \C; \quad (X, u) \mapsto \frac1T \int_0^T \exp\left(i u^\top X(t)\right) \, dt, \\
    \label{eq:tc}
    &\tc \colon \R^d \to \C; \quad u \mapsto \int_{\R^d} \exp\left(i u^\top x\right) \mu(\vt, x) \, dx.
\end{align}
For notational brevity, we will occasionally write $\ec{T}(X)$ to denote the function $\ec{T}(X, \cdot) \colon \R^d  \to \C$  whenever $X \in \mathscr{X}^T$ is fixed.

Let $L^2(\varphi) := L^2(\R^d, \mathscr{B}(\R^d), \varphi)$ be the weighted $L^2$-Hilbert space with some measurable Lebesgue-density $\varphi$ on $\R^d$. Define the distance quantity
\begin{equation} \label{eq:MDE_distance_measure}
    \dist(\vt, X) := \normLebesgue{\ec{T}(X) - \tc}{\varphi}^2 = \int_{\R^d} \left| \ec{T}(X, u) - \tc(u) \right|^2 \varphi(u) \, du, \quad \vt \in \Theta, \; X \in\mathscr{X}^T.
\end{equation}
We define the (plug-in) minimum distance estimator $\hvt_T$ (MDE) by means of the minimization task
\begin{equation} \label{eq:MDE}
    \mde{X} := \arginf_{\vt \in \Theta} \dist(\vt, X).
\end{equation}
Observe that $\tc$ is merely the characteristic function of the measure $\mu \, d\lambda^d$ and $\ec{T}$ is the empirical counterpart, where $\lambda^d$ denotes the $d$-dimensional Lebesgue measure. These kinds of estimator are well known in the statistics literature. For example, in \cite{K:1994} several minimum distance estimation approaches are rigorously discussed and studied for diffusion processes with small noise. Furthermore, questions of (asymptotic) existence and uniqueness of minimum distance estimators have been addressed and answered under fairly mild assumptions; see \cite{M:1984, K:1994}. We will not dwell on these assumptions for the MDE at this stage. Rather, we want to provide a specific example that will become important in the numerical experiments reported later in the paper.
\begin{Example} \label{ex:general_Gaussian_dist_exmaple}
Let $\vt\in \Theta$. Choose the weight $\varphi$ in \eqref{eq:MDE_distance_measure} as the density of a centered multivariate normal distribution with covariance matrix $\beta^2 I_d$, $\beta>0$, $I_d \in \R^{d \times d}$ the identity matrix, that is, 
\begin{equation}    \label{gaussian_weight}
    \varphi(u) = \frac{1}{\left(2\pi \beta^2\right)^{d/2}} \exp\left(-\frac{\left| u \right|_2^2}{2 \beta^2} \right), \quad u \in \R^d.
\end{equation}
The characteristic function of $\varphi$ is $k(x) :=\exp(-\beta^2 |x|_2^2/2)$, $x \in \R^d$, and a short calculation gives us the following formula for the distance quantity introduced in \eqref{eq:MDE_distance_measure}
\begin{align} \label{eq:distance_formula_gaussian_weight}
\begin{aligned}
    \dist(\vt, X) = 
    &\frac{1}{T^2} \int_0^T \int_0^T k(X(t)-X(s)) \, dt \, ds \\
    &- \frac{2}{T} \int_0^T (\mu(\vt) \ast k)(X(t)) \, dt + \int_{\R^d} (\mu(\vt) \ast k)(x) \mu(\vt, x) \, dx,
\end{aligned}
\end{align}
where $\ast$ denotes the convolution operator on $\R^d$. Going a step further, if $\mu(\vt)$ is the density of a centered multivariate normal distribution with a positive definite covariance matrix $\Sigma(\vt) \in \R^{d \times d}$, then formula \eqref{eq:distance_formula_gaussian_weight} simplifies to
\begin{align} \label{eq:distance_formula_normal_distr}
\begin{aligned}
    \dist(\vt, X) = 
    &\frac{1}{T^2} \int_0^T \int_0^T k(X(t)-X(s)) \, dt \, ds \\
    &- \frac{2}{T \sqrt{\det(I_d + \beta^2 \Sigma(\vt)) }} \int_0^T \exp\left( -\frac{\beta^2}{2} X(t)^\top \left( I_d + \beta^2 \Sigma(\vt) \right)^{-1} X(t) \right) \, dt \\
    &+ \frac{1}{\sqrt{\det(I_d + 2 \beta^2 \Sigma(\vt))}},
\end{aligned}
\end{align}
where $\det(\cdot)$ is the determinant.
\end{Example}

The example demonstrates that even in the simple case of a multivariate Gaussian case, it is impossible to derive a closed-form expression for the MDE, which, from a numerical point of view, necessitates the use of optimization algorithms. However, in a multivariate Gaussian setting, the formula \eqref{eq:distance_formula_normal_distr} makes the computation relatively tractable, which need not be the case for minimum distance estimators in general, cf. \cite{CM:2011}.

The setting we are about to study is fairly non-classical. To explain this, fix $\vt_0 \in \Theta$ and refer to this as the true parameter and to $\mathcal{M}_{\vt_0}$ as the true model. In the classical statistics literature we would assume that the data $X$ comes from the true model $\mathcal{M}_{\vt_0}$, is plugged into an estimator, e.g., an MDE, and the estimator yields, at least asymptotically, a close-to-correct estimate of the true parameter. However, in this paper, we want to allow for a specific misspecification between the true model and the model that generated the data. We will plug in the weakly perturbed data $X_\eps$, which will belong to another $\eps$-dependent model, into the MDE $\hvt_T$ and study the asymptotic properties of $\hvt_T(X_\eps)$ as $T \rightarrow \infty$ and $\epstozero$. Due to the assumed weak convergence of $X_\eps$ to $X$, it is desirable to have robustness guarantees for the MDE under such weak perturbations of $X$. This puts us directly into a setting of parameter estimation for misspecified models, cf. \cite{K:2004}.

\section{Asymptotic properties} \label{sec:asymptotic_properties}
We first extend some basic terminology that is commonly used in asymptotic estimation theory, cf. \cite{IK:1981}. Let $\Theta, \Tilde{\Theta} \subseteq \R^p$, $p \in \N$, be open, bounded sets and $\psi \colon \Theta \to \Tilde{\Theta}$ bijective. We consider the stochastic processes $X, X_\eps \in \mathscr{X}^T$ and assume that their laws with respect to $\Pb$ can be parameterized through a finite-dimensional parameter $\vt \in \Theta$, i.e., there exist two sets of probability measures on $(\mathscr{X}^{T}, \mathscr{B}^{T})$, say
\begin{equation}
    \mathscr{P} := \left\{ \Pb_\vt^X \, | \, \vt \in \Theta \right\} \quad \text{and} \quad \mathscr{P}_\eps := \left\{ \Pb_{\psi(\vt)}^{X_\eps} \, | \, \vt \in \Theta \right\},
\end{equation}
serving as distributional models for $X$ and $X_\eps$, respectively. Here, $\mathscr{B}^{T}$ is a $\sigma$-algebra on $\mathscr{X}^T$. In our specific context of weak perturbations, we will refer to $\mathscr{P}$ as limit model or effective model and to $\mathscr{P}_\eps$ as perturbed model. The function $\psi$, which is independent of $\eps$, draws a connection between the parameters of the two distributional models. We assume that the maps $\vt \mapsto \Pb_\vt^X$ and $\vt \mapsto \Pb_{\psi(\vt)}^{X_\eps}$ are bijective. If we, for a second, interpret $X_\eps$ as a realization of a random experiment on $(\mathscr{X}^T, \mathscr{B}^T)$, then the statistical task is to find the parameter $\vt_0 \in \Theta$ which gives the distributional model $\Pb_{\psi(\vt_0)}^{X_\eps}$ and, thus, also $\Pb_{\vt_0}^X$. We will call such a $\vt_0$ the true parameter. Finding the true parameter can be accomplished with a parametric estimator $\hat{\theta}_T \colon (\mathscr{X}^{T}, \mathscr{B}^{T}) \to (\Theta, \mathscr{B}(\R^p) \cap \Theta )$ for $\vt_0$.

In our particular setting of parameter estimation for misspecified models, we will either consider the following family of $T$-dependent parametric models
\begin{equation}
    \mathcal{M}^{T} := \left(\mathscr{X}^{T}, \mathscr{B}^{T},  \left\{ \Pb_{\psi(\vt)}^{X_\eps} \, | \, \vt \in \Theta \right\} \right), \quad T > 0,
\end{equation}
or the family of $\eps$-dependent parametric models
\begin{equation}
    \mathcal{M}^{T_\eps} := \left( \mathscr{X}^{T_\eps}, \mathscr{B}^{T_\eps},  \left\{ \Pb_{\psi(\vt)}^{X_\eps} \, | \, \vt \in \Theta \right\} \right), \quad \eps > 0,
\end{equation}
where $T_\eps>0$ depends on the parameter $\eps$. Note also that in the classical parameter estimation setting for specified models, we would rather work with the family of parametric models
\begin{equation*}
    \left(\mathscr{X}^{T}, \mathscr{B}^{T},  \left\{ \Pb_{\vt}^{X} \, | \, \vt \in \Theta \right\} \right), \quad T > 0.
\end{equation*}

\begin{Example}
Consider the following multiscale overdamped Langevin stochastic differential equation (SDE) in one dimension with $\alpha, \sigma > 0$
\begin{equation} \label{eq:multiscale_langevin_linear_exa}
    dX_\eps(t) = -\alpha X_\eps(t) - \frac{1}{\eps} \cos\left( \frac{2 \pi X_\eps(t)}{\eps} \right) dt  + \sqrt{2 \sigma} dW(t), \quad t \in [0,T].
\end{equation}
By homogenization theory, see \cite[Chapter 3]{BLP:1978}, $\Pb^{X_\eps}$ converges weakly to $\Pb^{X}$ in $C([0,T];\R^d)$ as $\epstozero$, where $X$ is the solution of the SDE
\begin{equation} \label{eq:limit_langevin_linear_exa}
    dX(t) = -\alpha K X(t) dt  + \sqrt{2 \sigma K} dW(t),  \quad t \in [0,T].
\end{equation}
Here, the constant factor $K>0$ emerges from the cell problem of the homogenization and is given by
\begin{equation}
    K := \frac{1}{Z^+ Z^-}, \quad Z^{\pm} := \int_0^{1} \exp \left( \pm \frac{\sin(2 \pi y)}{\sigma} \right) \, dy.
\end{equation}
If the statistical task is the estimation of the drift parameter $\vt := \alpha K \in (0, \infty)$ in \eqref{eq:limit_langevin_linear_exa}, then we may consider the $\eps$-dependent model
\begin{equation}
    \mathcal{M}^{T_\eps} := \left(C([0,T_\eps];\R^d), \mathscr{B}^{T_\eps}, \left\{ \Pb_{\psi(\vt)}^{X_\eps} \, | \, \vartheta \in \Theta_0 \right\} \right), \quad \eps > 0,
\end{equation}
where $\mathscr{B}^{T}$ is a $\sigma$-algebra on $C([0,T_\eps];\R^d)$, $T_\eps \rightarrow \infty$ as $\epstozero$, and $\Theta_0 \subset (0, \infty)$ is an open, bounded interval. The function $\psi$ is given, in this case, by $\psi(\vt) = \vt/K$. We will return to a more general version of this example in Section \ref{sec:langevin}.
\end{Example}

In order to quantify properties of estimators, we will say that the family of estimators 
\begin{equation}
    \hat{\theta}_T \colon (\mathscr{X}^{T}, \mathscr{B}^{T}) \to (\Theta, \mathscr{B}(\R^p) \cap \Theta ), \quad \; T > 0,
\end{equation}
is $\eps$-$T$-robust ($T$-$\eps$-robust) for the true parameter $\vt_0$ if for all $\delta > 0$
\begin{equation}
    \lim_{\epstozero} \lim_{T \rightarrow \infty} \spm{\psi(\vt_0)}{}( |\hat{\theta}_{T}(X_\eps) - \vt_0 |_2 > \delta ) = 0, \quad
    \left( \lim_{T \rightarrow \infty} \lim_{\epstozero} \spm{\psi(\vt_0)}{}( |\hat{\theta}_{T}(X_\eps) - \vt_0 |_2 > \delta ) = 0 \right).
\end{equation}
We will say that the family of estimators is 
\begin{equation}
    \hat{\theta}_{T_\eps} \colon (\mathscr{X}^{T_\eps}, \mathscr{B}^{T_\eps}) \to (\Theta, \mathscr{B}(\R^p) \cap \Theta ), \quad \; \eps > 0,
\end{equation}
is $\eps$-robust for the true parameter $\vt_0$ if for all $\delta > 0$
\begin{equation}
    \lim_{\epstozero} \spm{\psi(\vt_0)}{}( |\hat{\theta}_{T_\eps}(X_\eps) - \vt_0 |_2 > \delta ) = 0.
\end{equation}
Observe that these definitions bear resemblance with the concept of consistency of estimators, cf. \cite{K:2004}; however, since we do not assume that the given observations are generated from a distribution in $\mathscr{P}$, they have been adjusted accordingly. Obtaining estimators with these robustness guarantees is, in fact, not a trivial task as it has proved shown for the classical maximum likelihood estimator (MLE) in a multiscale SDE setting, see \cite{PS:2007, PPS:2009}. In these papers it has been proved that the classical MLE is not robust under multiscale observations when the MLE is built on the basis of the homogenized limit equation.

Suppose $\vt_0 \in \Theta$ is the true parameter, and that there is another density $\mu_\eps(\psi(\vt_0))$ associated to the process $X_\eps$ for which \eqref{eq:ergodicity_abstract} holds for every fixed $\eps > 0$. Let $\xi, \xi_\eps \colon \Omega \to \R^d$ be random vectors with $\xi \sim \mu(\vt_0, \cdot) \, d \lambda^d$ and $\xi_\eps \sim \mu_\eps(\psi(\vt_0), \cdot) \, d \lambda^d$. We abbreviate $\mu = \mu(\vt_0) = \mu(\vt_0, \cdot)$ and $\mu_\eps = \mu_\eps(\psi(\vt_0)) = \mu_\eps(\psi(\vt_0), \cdot)$ whenever $\vt_0 \in \Theta$ is fixed as the true parameter.

\begin{Proposition} \label{prop:robustness_eps_T}
Fix $\vt_0 \in \Theta$ as the true parameter. Assume that $\xi_\eps$ converges weakly to $\xi$ in $\R^d$ as $\epstozero$, and that $X_\eps$ satisfies a mean ergodic theorem on $C_b(\R^d; \C)$ for all $\eps > 0$, i.e., for any $h \in C_b(\R^d; \C)$ and any $\eps > 0$ it holds 
\begin{equation} \label{eq:L^2-ergodic}
    \E_{\psi(\vt_0)} \Bigg| \frac1T \int_0^T h(X_\eps(t)) \, dt - \int_{\R^d} h(x)  \mu_\eps( x) \, dx \Bigg|^2  \rightarrow 0, \quad \Ttoinfty.
\end{equation}
Suppose further that the identifiability condition is satisfied, that is, for all $\delta > 0$
\begin{equation}    \label{eq:identifiability_cond}
    d_\delta(\vt_0) := \inf_{|\vt - \vt_0|_2 > \delta} \normLebesgue{\tc - \mathscr{C}_{\vt_0}}{\varphi} > 0.
\end{equation}
Then the family of estimators $(\hvt_T)$ is $\eps$-$T$-robust for $\vt_0$.
\end{Proposition}
\begin{proof}
First, we note that, by the definition of MDE, for all $\delta > 0$
\begin{equation*}
    \left\{ \; |\mde{X_\eps} - \vt_0|_2 > \delta \; \right\} 
    = \left\{ \; \inf_{|\vt - \vt_0|_2\leq \delta} \normLebesgue{\ec{T}(X_\eps) - \tc}{\varphi} > \inf_{|\vt - \vt_0|_2> \delta} \normLebesgue{\ec{T}(X_\eps) - \tc}{\varphi} \; \right\}.
\end{equation*}
It follows that
\begin{align*}
    \spm{\psi(\vt_0)}{}( |\mde{X_\eps} - \vt_0|_2 > \delta ) 
    &= \spm{\psi(\vt_0)}{}\left( \inf_{|\vt - \vt_0|_2 \leq \delta} \normLebesgue{\ec{T}(X_\eps) - \tc}{\varphi} > \inf_{|\vt - \vt_0|_2 > \delta} \normLebesgue{\ec{T}(X_\eps) - \tc}{\varphi} \right) \\
    &\leq \spm{\psi(\vt_0)}{}\Bigg( \inf_{|\vt - \vt_0|_2 \leq \delta} \normLebesgue{\ec{T}(X_\eps) - \mathscr{C}_{\vt_0}}{\varphi} \\
    & \hspace{1.25cm} + \inf_{|\vt - \vt_0|_2 \leq \delta} \normLebesgue{\tc - \mathscr{C}_{\vt_0}}{\varphi} > \inf_{|\vt - \vt_0|_2 > \delta} \normLebesgue{\ec{T}(X_\eps) - \tc}{\varphi} \Bigg).
\end{align*}
We use the fact that $\inf_{|\vt - \vt_0|_2 \leq \delta} \normLebesgue{\tc - \mathscr{C}_{\vt_0}}{\varphi} = 0$, the triangle inequality, and Markov's inequality to deduce
\begin{align*}
    \spm{\psi(\vt_0)}{}( |\mde{X_\eps} - \vt_0|_2 > \delta ) 
    &\leq \spm{\psi(\vt_0)}{} \Bigg( \normLebesgue{\ec{T}(X_\eps) - \mathscr{C}_{\vt_0}}{\varphi} > \inf_{|\vt - \vt_0|_2 > \delta} \normLebesgue{\tc - \mathscr{C}_{\vt_0}}{\varphi} \\
    & \hspace{1.25cm} - \normLebesgue{\ec{T}(X_\eps) - \mathscr{C}_{\vt_0}}{\varphi} \Bigg) \\
    &= \spm{\psi(\vt_0)}{} \Bigg( 2 \normLebesgue{\ec{T}(X_\eps) - \mathscr{C}_{\vt_0}}{\varphi} > d_\delta(\vt_0) \Bigg) \\
    &\leq \frac{4}{d_\delta(\vt_0)^2} \E_{\psi(\vt_0)} \dist(\vt_0, X_\eps)\,.
\end{align*}
Next, we define
\begin{equation*}
    \mathscr{C}_\eps(u) := \int_{\R^d} \exp \left( i u^\top x \right)  \mu_\eps(x) \, dx, \quad u \in \R^d,
\end{equation*}
and bound the last expectation by
\begingroup
\allowdisplaybreaks
\begin{align*}
    \E_{\psi(\vt_0)} \dist(\vt_0, X_\eps) 
    \leq 2 \left[ \E_{\psi(\vt_0)} \normLebesgue{\ec{T}(X_\eps) - \mathscr{C}_\eps}{\varphi}^2 + \normLebesgue{\mathscr{C}_\eps - \mathscr{C}_{\vt_0}}{\varphi}^2 \right]\,.
\end{align*}
\endgroup
By assumption, we have, for the first term,
\begin{equation*}
    \E_{\psi(\vt_0)} \normLebesgue{\ec{T}(X_\eps) - \mathscr{C}_\eps}{\varphi}^2 = \int_{\R^d} \E_{\psi(\vt_0)} \Bigg| \frac1T \int_0^T \exp \left( i u^\top X_\eps(t) \right) \, dt - \mathscr{C}_\eps(u) \Bigg|^2 \varphi(u) \, du
    \rightarrow 0, \quad \Ttoinfty,
\end{equation*}
and, for the second term, we have that $\xi_\eps$ converges weakly to $\xi$ in $\R^d$ as $\epstozero$, so that, by Lévy's continuity theorem, it follows that $\mathscr{C}_\eps \rightarrow \mathscr{C}_{\vt_0}$ pointwise on $\R^d$ as $\epstozero$. An application of the dominated convergence theorem gives us
\begin{equation*}
    \normLebesgue{\mathscr{C}_\eps - \mathscr{C}_{\vt_0}}{\varphi} \rightarrow 0, \quad \epstozero. 
\end{equation*}
Putting everything together, we obtain
\begin{align*}
    \lim_{\epstozero} \lim_{T \rightarrow \infty} \spm{\psi(\vt_0)}{}( |\mde{X_\eps} - \vt_0| > \delta ) = 0.
\end{align*}
\end{proof}
\begin{Remark}  \label{rem:remark_L^2_ergodic}
If we exchange the condition \eqref{eq:L^2-ergodic} by the condition that for  any $h \in C_b(\R^d; \C)$ it holds that
\begin{equation} \label{eq:eps_L^2-ergodic}
    \E_{\psi(\vt_0)} \Bigg| \frac{1}{T_\eps} \int_0^{T_\eps} h(X_\eps(t)) \, dt - \int_{\R^d} h(x)  \mu_\eps( x) \, dx \Bigg|^2  \rightarrow 0, \quad \epstozero,
\end{equation}
then, by the same proof, we obtain the $\eps$-robustness of $(\hvt_{T_\eps})$. A result of this kind can be found in \cite{BK:2025}.
\end{Remark}

In the next proposition, we exchange the order of the limits and prove the $T$-$\eps$-robustness under different assumptions. We briefly want to recall that the space $C_b(\R^d; \C)$ is equipped with the metric
\begin{equation}    \label{eq:metric_C(R^d)}
    \rho(f,g) := \sum_{n=1}^\infty \frac{1}{2^n} \frac{\rho_n(f,g)}{1+\rho_n(f,g)} , \quad \rho_n(f,g) := \sup_{x \in [-n,n]^d} |f(x) - g(x)|, \quad f, g \in C(\R^d; \C),
\end{equation}
which makes $C(\R^d; \C)$ a complete, separable metric space. For details on tightness and weak convergence in $C(\R^d; \C)$, we refer the reader to \cite[Section 2.4]{KS:1996}, where the results are established for the space $C([0,\infty);\R)$, but they can be formulated and proved with only minor modifications for $C(\R^d; \C)$, as well.
\begin{Proposition} \label{prop:robustness_T_eps}
Suppose $\mathscr{X}^T \hookrightarrow L^1([0, T], \R^d)$. Let $\vt_0 \in \Theta$, $d_\delta(\vt_0) > 0$, and $X$ satisfy an ergodic theorem on $C_b(\R^d; \C)$, either in the mean or almost sure sense. Under the assumptions that $\E_{\psi(\vt_0)} \left| \xi \right|_2^2 < \infty$ and
\begin{equation}    \label{eq:moment_bounds_eps}
    \sup_{\eps > 0} \int_0^T \E_{\psi(\vt_0)} \left| X_\eps(t) \right|_2^2 \, dt \leq C_T,
\end{equation}
for some constant $C_T > 0$ that can depend on $T$, the family of estimators $(\hvt_T)$ is $T$-$\eps$-robust for $\vt_0$.
\end{Proposition}

\begin{proof}
Fix $\vt_0 \in \Theta$. We established in the preceding proof that
\begin{equation} \label{eq:main_ineq}
    \spm{\psi(\vt_0)}{}( |\mde{X_\eps} - \vt_0|_2 \geq \delta ) \leq \spm{\psi(\vt_0)}{} \Bigg( 2 \normLebesgue{\ec{T}(X_\eps) - \mathscr{C}_{\vt_0}}{\varphi} \geq d_\delta(\vt_0) \Bigg)
\end{equation}    
for any $\eps > 0$ and $T > 0$. The last expression motivates the definition of the map
\begin{equation}
    Z_T(X_\eps, u) := \ec{T}(X_\eps, u) - \mathscr{C}_{\vt_0}(u) = \frac1T \int_0^T \exp \left( i u^\top X_\eps(t) \right) - \E_{\psi(\vt_0)} \exp \left( i u^\top \xi \right) \, dt, \quad u \in \R^d,
\end{equation}
with $T>0$ fixed. The function $Z_T(X_\eps, \cdot) \colon \R^d \to \C$ can be interpreted as a random element of $C(\R^d; \C)$ and the functional $Z_T(\cdot , u) \colon (\mathscr{X}^T, \| \cdot \|_{\mathscr{X}^T}) \to \C$ with fixed $u \in \R^d$ is bounded and continuous. Indeed, for fixed $u \in \R^d$ and any $f, g  \in \mathscr{X}^T$ we have $|Z_T(f, u)| \leq 2$ and
\begin{align*}
    |Z_T(f, u) - Z_T(g, u)| &= \Bigg| \frac1T \int_0^T \exp \left( i u^\top f_t \right) - \exp \left( i u^\top g_t \right) \, dt \Bigg| \leq \frac1T \int_0^T \Bigg| \int_{u^\top g_t}^{u^\top f_t} i \exp \left( is \right) \, ds \Bigg| \, dt \\
    &\leq \frac1T \int_0^T  \left| u^\top (f_t - g_t) \right| \, dt \leq \frac{C_e}{T} |u|_2  \| f - g \|_{\mathscr{X}^T}\,,
\end{align*}
where $C_e > 0$ is the constant coming from the continuous embedding $\mathscr{X}^T \hookrightarrow L^1([0, T], \R^d)$. Hence, $Z_T(\cdot, u)$ is even Lipschitz-continuous.
In the following we prove for $T > 0$
\begin{equation} \label{eq:Z_T_weak_conv}
    Z_T(X_\eps, \cdot) \weakconvunder{\psi(\vt_0)} Z_T(X, \cdot) \quad \text{in } C(\R^d; \C) \text{ as } \epstozero.
\end{equation}
If not stated otherwise, the subsequent limit behavior is always understood as $\epstozero$. We fix arbitrary vectors $u_1, \ldots, u_k \in \R^d$, $k \in \N$, and define the map
\begin{equation*} 
    F \colon (\mathscr{X}^T, \| \cdot \|_{\mathscr{X}^T}) \to (\C^k, \| \cdot \|_{\C^k}) ; \quad f \mapsto (Z_T(f, u_1), \ldots, Z_T(f, u_k)).
\end{equation*}
It immediately follows by the calculation before that $F$ is Lipschitz-continuous:
\begin{equation*}
    \| F(f) - F(g) \|_{\C^k} \leq \frac{C_e}{T} \Bigg( \sum_{i=1}^k | u_i |_2\Bigg )^{1/2} \| f-g \|_{\mathscr{X}^T}, \quad f, g \in \mathscr{X}^T.
\end{equation*}
The weak convergence of $X_\eps$ in $\mathscr{X}^T$ and the continuous mapping theorem yield the weak convergence of the finite-dimensional distributions of $Z_T(X_\eps, \cdot)$:
\begin{equation*}
    (Z_T(X_\eps, u_1), \ldots Z_T(X_\eps, u_k)) \weakconvunder{\psi(\vt_0)} (Z_T(X, u_1), \ldots Z_T(X, u_k)) \quad \text{in } \C^k.
\end{equation*}
To finish the proof of \eqref{eq:Z_T_weak_conv}, it suffices to establish the tightness of $(Z_T(X_\eps, \cdot))_{\eps > 0}$ in the space $C(\R^d; \C)$ as the weak convergence then follows from \cite[Theorem 2.4.15]{KS:1996}. For this purpose, we aim at establishing the two conditions of \cite[Problem 2.4.11]{KS:1996}, which require that $(Z_T(X_\eps, \cdot))_{\eps > 0}$ satisfies for some positive constants $\alpha, \beta, \nu > 0$
\begin{equation}    \label{eq:tightness_cond1}
    \sup_{\eps > 0} \E_{\psi(\vt_0)} \left| Z_T(X_\eps, 0) \right|^\nu < \infty,
\end{equation}
and for all $k \in \N$ and $u, v \in \R_{\leq k}^d := \{x \in \R^d \, | \, |x|_2 \leq k \}$
\begin{equation}    \label{eq:tightness_cond2}
    \sup_{\eps > 0} \E_{\psi(\vt_0)} \Big| Z_T(X_\eps, u) - Z_T(X_\eps, v) \Big|_2^\alpha \leq C_k \left| u - v \right|_2^{1 + \beta},
\end{equation}
where $C_k > 0$ can depend on $k$. We will establish these conditions for $\alpha = \nu = 2$ and $\beta = 1$. The first condition \eqref{eq:tightness_cond1} is obvious, since $Z_T(X_\eps, 0) = 0$ for all $\eps > 0$. For the second condition \eqref{eq:tightness_cond2}, we estimate
\begingroup
\allowdisplaybreaks
\begin{align*}
    \E_{\psi(\vt_0)} \Big| Z_T(X_\eps, u) - Z_T(X_\eps, v) \Big|_2^2 &\leq 2 \Bigg[ \E_{\psi(\vt_0)} \left| \frac1T \int_0^T \exp \left( i u^\top X_\eps(t) \right) -  \exp \left( i v^\top X_\eps(t) \right) \, dt \right|^2 \\
    &\hspace{0.5cm}+ \left| \mathscr{C}_{\vt_0}(u) - \mathscr{C}_{\vt_0}(v) \right|^2 \Bigg] \\
    &\leq 2 \left[ \E_{\psi(\vt_0)} \left( \frac1T \int_0^T \left| X_\eps(t) \right|_2 \, dt \right)^2 |u-v|_2^2 + \E_{\psi(\vt_0)} \left| \xi \right|_2^2 \left| u - v \right|_2^2 \right] \\
    &\leq 2 \left[ \E_{\psi(\vt_0)} \left( \frac1T \int_0^T \left| X_\eps(t) \right|_2 \, dt \right)^2 + \E_{\psi(\vt_0)} \left| \xi \right|_2^2 \right] \left| u - v \right|_2^2 \\
    &\leq 2 \left[ \frac1T \int_0^T \E_{\psi(\vt_0)} \left| X_\eps(t) \right|_2^2 \, dt + \E_{\psi(\vt_0)} \left| \xi \right|_2^2 \right] \left| u - v \right|_2^2,
\end{align*}
\endgroup
so that, using the assumptions, \eqref{eq:tightness_cond2} is satisfied, as well, and \eqref{eq:Z_T_weak_conv} follows. Observe that $Z_T(X, \cdot) \in C_b(\R^d; \C)$ $\Pb_{\psi(\vt_0)}$-a.s., so that the numerical function $\normLebesgue{\cdot}{\varphi} \colon C(\R^d; \C) \to [0, \infty]$ is continuous $\Pb^{Z_T(X, \cdot)}_{\psi(\vt_0)}$-a.s., but it is also measurable. Indeed, the measurability follows from the fact that the sequence of functions $\Phi_n \colon (C(\R^d; \C), \rho) \to \R$, $n \in \N$, defined by
\begin{equation*}
    \Phi_n(f) := \int_{[-n,n]^d} |f(x)|^2 \varphi(x) \, dx, \quad f \in C(\R^d; \C),
\end{equation*}
are continuous. This claim can be seen from the well-known fact that for $f_k, f \in C(\R^d; \C)$
\begin{equation*}
    \rho(f_k,f) = \sum_{n=1}^\infty \frac{1}{2^n} \frac{\rho_n(f_k,f)}{1+\rho_n(f_k,f)} \rightarrow 0 \quad \Leftrightarrow \quad \rho_n(f_k,f) \rightarrow 0 \quad \forall n \in \N, \quad \textrm{as } k \rightarrow \infty.
\end{equation*}
Consequently, if $n \in \N$ and $f \in C(\R^d; \C)$ are fixed and $(f_k)_{k \in \N} \subset C(\R^d; \C)$ is a sequence such that $f_k \rightarrow f$ as $k \rightarrow \infty$, then 
\begin{equation*}
    \left| \Phi_n(f_k) - \Phi_n(f) \right| \leq \rho_n(f_k, f) \left( \rho_n(f_k, f) + 2 \sup_{x \in [-n,n]^d} |f(x)|  \right) \rightarrow 0, \quad k \rightarrow \infty.
\end{equation*}
In particular, the functions $\Phi_n$ are measurable. By the monotone convergence theorem, they converge pointwise on $C(\R^d; \C)$ to $\normLebesgue{\cdot}{\varphi}$. Therefore, a variant of the continuous mapping theorem, see \cite[2.2.9 Theorem]{B:2018}, yields
\begin{equation*}
    \normLebesgue{\ec{T}(X_\eps) - \mathscr{C}_{\vt_0}}{\varphi} \weakconvunder{\psi(\vt_0)} \normLebesgue{\ec{T}(X) - \mathscr{C}_{\vt_0}}{\varphi} \quad \text{in } \R.
\end{equation*}
Returning to \eqref{eq:main_ineq}, we therefore obtain
\begin{equation*}
    \limsup_{\epstozero} \spm{\psi(\vt_0)}{} \Bigg( 2 \normLebesgue{\ec{T}(X_\eps) - \mathscr{C}_{\vt_0}}{\varphi} \geq d_\delta(\vt_0) \Bigg) \leq \spm{\psi(\vt_0)}{} \Bigg( 2 \normLebesgue{\ec{T}(X) - \mathscr{C}_{\vt_0}}{\varphi} \geq d_\delta(\vt_0) \Bigg).
\end{equation*}
The claim now follows from the ergodic theorem on $C_b(\R^d; \C)$ for $X$. 
\end{proof}

Our next result concerns a representation of the difference $\sqrt{T_\eps}(\hvt_{T_\eps}(X_\eps) - \vt_0)$, $\vt_0 \in \Theta$, in a way that eventually allows for a proof of the asymptotic normality of the MDE as $\epstozero$.
\begin{Proposition} \label{prop:asymptotic_normality}
    Let $\vt_0 \in \Theta$, where $\Theta$ is an open, bounded, convex set in $\R^p$. Assume that the function $\vt \mapsto \tc(u)$, $u \in \R^d$, is twice continuously differentiable with respect to $\vt \in \Theta$. We denote the first derivative by $\partial_\vt \tc(u)$, $u \in \R^d$, and interpret it as a column vector in $\C^p$. Assume further that 
    \begin{equation}
        \left|\partial_\vt \tc(u)\right|_2 \leq B^{(1)}(u), \quad \left\|\partial^2_\vt \, \tc(u)\right\|_F \leq B^{(2)}(u), \quad \vt \in \Theta,
    \end{equation}
    with $B^{(1)}, B^{(2)} \in L^1(\varphi) := L^1(\R^d, \mathscr{B}(\R^d), \varphi)$. If $(\hvt_{T_\eps})$ is $\eps$-robust for the true parameter $\vt_0$, and the matrix
    \begin{equation} \label{eq:limit_variance}
        J(\vt_0) := \int_{\R^d} \partial_\vt \mathscr{C}_{\vt_0}(u) \, \partial_\vt \mathscr{C}_{\vt_0}(u)^\top \varphi(u) \, du
    \end{equation}
    is positive definite, then as $\epstozero$
    \begin{equation} \label{eq:asymptotic_normality}
    \sqrt{T_\eps} (\hvt_{T_\eps}(X_\eps) - \vt_0) = \frac{J(\vt_0)^{-1}}{\sqrt{T_\eps}} \int_0^{T_\eps} \int_{\R^d} \left( \exp \left( i u^\top X_\eps(t) \right) - \E \exp \left( i u^\top \xi \right) \right) \, \partial_\vt \mathscr{C}_{\vt_0}(u) \varphi(u) \, du \, dt + o_{\, \Pb^{\, \eps}_{\vt_0}}(1),
    \end{equation}
    where we put $\Pb^{\, \eps}_{\vt_0} := \Pb_{\psi(\vt_0)}^{\hat{\vt}_{T_\eps}(X_\eps)}$.
\end{Proposition}

\begin{proof}
Fix $\vt_0 \in \Theta$ and write, for the sake of readability, $\hvt := \hvt_{T_\eps}(X_\eps)$ in the following. Differentiating \eqref{eq:MDE_distance_measure} with respect to $\vartheta$ and evaluating the resulting equation at $\hvt$ yields
\begin{equation} \label{eq:minimum_distance_equation}
    \int_{\R^d} \left( \ec{T_\eps}(X_\eps, u) - \mathscr{C}_{\hvt}(u) \right) \, \partial_\vt \mathscr{C}_{\hvt}(u) \varphi(u) \, du = 0.
\end{equation}
A first-order Taylor expansion for multivariate functions, see \cite[Theorem B.6.1]{S:2013}, gives us for all $u \in \R^d$
\begin{equation} \label{eq:taylor_expansion}
    \mathscr{C}_{\hvt}(u) - \mathscr{C}_{\vt_0}(u) = \partial_\vt \mathscr{C}_{\vt_0}(u)^\top (\hvt - \vt_0) + R_{\hvt}(u)^\top (\hvt - \vt_0)\,,
\end{equation}
where
\begin{equation*}
    R_{\hvt}(u) = \int_0^1 \partial_\vt \mathscr{C}_{\vt_0 + r(\hvt - \vt_0)}(u) - \partial_\vt \mathscr{C}_{\vt_0}(u) \, dr.
\end{equation*}
The remainder term satisfies the inequality
\begin{equation*}  
    \left| R_{\hvt}(u) \right|_2 \leq \sup_{\vt \in S[\vt_0, \hvt]} \left| \partial_\vt \mathscr{C}_{\vt}(u)  - \partial_\vt \mathscr{C}_{\vt_0}(u) \right|_2, 
\end{equation*}
where $S[\vt_0, \hvt] \subseteq \Theta$ is the line segment that connects $\vt_0$ and $\hvt$. Inserting \eqref{eq:taylor_expansion} into \eqref{eq:minimum_distance_equation} gives
\begin{align*}
    0 &= \int_{\R^d} \left[ \ec{T_\eps}(X_\eps, u) - \mathscr{C}_{\vt_0}(u) - \left( \mathscr{C}_{\hvt}(u) - \mathscr{C}_{\vt_0}(u) \right) \right] \, \partial_\vt \mathscr{C}_{\hvt}(u) \varphi(u) \, du \\
    &= \int_{\R^d} \left[ \ec{T_\eps}(X_\eps, u) - \mathscr{C}_{\vt_0}(u) - \left( \partial_\vt \mathscr{C}_{\vt_0}(u)^\top (\hvt - \vt_0) + R_{\hvt}(u)^\top (\hvt - \vt_0) \right) \right] \, \partial_\vt \mathscr{C}_{\hvt}(u) \varphi(u) \, du,
\end{align*}
and rearranging yields
\begin{align*}
    J(\hvt, \vt_0) (\hvt - \vt_0) = \int_{\R^d} \left( \ec{T_\eps}(X_\eps, u) - \mathscr{C}_{\vt_0}(u) - R_{\hvt}(u)^\top (\hvt - \vt_0) \right) \, \partial_\vt \mathscr{C}_{\hvt}(u) \varphi(u) \, du,
\end{align*}
with
\begin{equation*}
    J(\hvt, \vt_0) := \int_{\R^d} \partial_\vt \mathscr{C}_{\hvt}(u) \, \partial_\vt \mathscr{C}_{\vt_0}(u)^\top \varphi(u) \, du.
\end{equation*}
Using another first-order Taylor expansion, the boundedness assumption on the derivatives, and the $\eps$-robustness gives, on the one hand,
\begin{align*}
    J(\hvt, \vt_0) = J(\vt_0) + o_{\, \Pb^{\, \eps}_{\vt_0}}(1), \quad \epstozero,
\end{align*}
and, on the other hand,
\begin{align*}
   &\int_{\R^d} \left( \ec{T_\eps}(X_\eps, u) - \mathscr{C}_{\vt_0}(u) - R_{\hvt}(u)^\top (\hvt - \vt_0) \right) \, \partial_\vt \mathscr{C}_{\hvt}(u) \varphi(u) \, du \\
   = &\int_{\R^d} \left( \ec{T_\eps}(X_\eps, u) - \mathscr{C}_{\vt_0}(u) \right) \, \partial_\vt \mathscr{C}_{\vt_0}(u) \varphi(u) \, du + o_{\, \Pb^{\, \eps}_{\vt_0}}(1), \quad \epstozero.
\end{align*}
It follows that, as $\epstozero$,
\begin{align*}
    \sqrt{T_\eps} (\hvt - \vt_0) 
    &= J(\vt_0)^{-1} \int_{\R^d} \sqrt{T_\eps} \left( \ec{T_\eps}(X_\eps, u) - \mathscr{C}_{\vt_0}(u) \right) \, \partial_\vt \mathscr{C}_{\vt_0}(u) \varphi(u) \, du + o_{\, \Pb^{\, \eps}_{\vt_0}}(1), \\
    &= J(\vt_0)^{-1} \frac{1}{\sqrt{T_\eps}} \int_0^{T_\eps} \int_{\R^d} \left( \exp \left( i u^\top X_\eps(t) \right) - \E \exp \left( i u^\top \xi \right) \right) \, \partial_\vt \mathscr{C}_{\vt_0}(u) \varphi(u) \, du \, dt + o_{\, \Pb^{\, \eps}_{\vt_0}}(1).
\end{align*}
\end{proof}

We define for $\eps > 0$
\begin{align} 
    \label{eq:asy_double_integral}
    I(\eps) &:= \frac{1}{\sqrt{T_\eps}} \int_0^{T_\eps} \int_{\R^d} \left( \exp \left( i u^\top X_\eps(t) \right) - \E \exp \left( i u^\top \xi \right) \right) \, \partial_\vt \mathscr{C}_{\vt_0}(u) \varphi(u) \, du \, dt \\
    \label{eq:asy_double_integral_split}
    &=: I_1(\eps) + I_2(\eps).
\end{align}
with
\begin{align}
    \label{eq:asy_double_integral_parts_1}
    &I_1(\eps) := \frac{1}{\sqrt{T_\eps}} \int_0^{T_\eps} \int_{\R^d} \left( \exp \left( i u^\top X_\eps(t) \right) - \E \exp \left( i u^\top \xi_\eps \right) \right) \, \partial_\vt \mathscr{C}_{\vt_0}(u) \varphi(u) \, du \, dt, \\
    \label{eq:asy_double_integral_parts_2}
    &I_2(\eps) := \sqrt{T_\eps} \int_{\R^d} \left( \E \exp \left( i u^\top \xi_\eps \right) - \E \exp \left( i u^\top \xi \right) \right) \, \partial_\vt \mathscr{C}_{\vt_0}(u) \varphi(u) \, du\,.
\end{align}
Inspecting the double integral in \eqref{eq:asy_double_integral_parts_1}, it is tempting to use a central limit theorem (CLT) to a normal distribution for the time integral
\begin{equation*}
    \frac{1}{\sqrt{T_\eps}} \int_0^{T_\eps} h_\eps(X_\eps(t)) \, dt,
\end{equation*}
where
\begin{equation}
    h_\eps(x) := \int_{\R^d} \left( \exp \left( i u^\top x \right) - \E \exp \left( i u^\top \xi_\eps \right) \right) \, \partial_\vt \mathscr{C}_{\vt_0}(u) \varphi(u) \, du, \quad x \in \R^d.
\end{equation}
While such a CLT would provide us with the asymptotic normality of the MDE, it cannot simply follow from our general setting. As we will see in the following section, this is by no means trivial and requires a careful investigation of the limit behavior of \eqref{eq:asy_double_integral_parts_1} and \eqref{eq:asy_double_integral_parts_2}. On the one hand, this is due to the circumstance that we need a good estimate for the appearing integral in \eqref{eq:asy_double_integral_parts_2}, so that we can leverage it against the $\sqrt{T_\eps}$ term, and, on the other and, the quantities in the time integral in \eqref{eq:asy_double_integral_parts_1} all depend on $\epsilon$, making the analysis significantly more difficult.

\section{Multiscale overdamped Langevin diffusion}    \label{sec:langevin}
In this section, we want to study a very specific class of multiscale diffusion processes with a homogenized limit process and apply the results of the preceding sections. Some of the required technical results can be found in Appendix \ref{app:A}, while others are contained in \cite{BK:2025}.

Consider the following multiscale overdamped Langevin SDE in one dimension
\begin{equation} \label{eq:multiscale_langevin}
    dX_\eps(t) = -\alpha V'(X_\eps(t)) - \frac{1}{\eps} p'\left( \frac{X_\eps(t)}{\eps} \right) dt  + \sqrt{2 \sigma} dW(t), \quad t \in [0,T],
\end{equation}
where $V \colon \R \rightarrow \R$, $\alpha, \sigma, \eps > 0$, and $p \in C^2([0,1]; \R)$ is a 1-periodic function. Fix $k \in \N$ for the rest of the section. We assume that $V \in C^k(\R; \R)$ with $V(0)=0$ has a global lower bound, grows at most polynomially, and there exist constants $a, b > 0$ such that
\begin{equation}    \label{eq:ergodicity_condition}
    -V'(x)x \leq a - b x^2, \quad x \in \R,
\end{equation}
so that $V$ is a so-called potential according to Definition \ref{def:potential} from the appendix. \\ \indent 
Using homogenization theory, see Chapter 3 of \cite{BLP:1978}, it can be shown that $\Pb^{X_\eps}$ converges weakly to $\Pb^{X}$ in $C([0,T];\R)$ as $\epstozero$, where $X$ is the solution of the SDE
\begin{equation} \label{eq:limit_langevin}
    dX(t) = -\alpha K V'(X(t)) dt  + \sqrt{2 \sigma K} dW(t),  \quad t \in [0,T].
\end{equation}
The constant factor $K>0$ emerges from the cell problem of the homogenization of \eqref{eq:multiscale_langevin} and equals
\begin{equation}
    K := \frac{1}{Z^+ Z^-}, \quad Z^{\pm} := \int_0^{1} \exp \left( \pm \frac{p(y)}{\sigma} \right) \, dy.
\end{equation}
Under the standing assumptions, it is not difficult to prove the validity of Assumptions (C) and (MET) of \cite{BK:2025}, so that the invariant densities $\mu_\eps$ and $\mu$ of $X_\eps$ and $X$, respectively, exist and are given by
\begin{equation}    \label{eq:eps_invariant_density_langevin}
    \mu_\eps(x) = \frac{1}{Z_\eps} \exp \left( -\frac{\alpha}{\sigma} V(x) - \frac{1}{\sigma} p\left(x/\eps\right) \right), \quad \mu(x) = \frac{1}{Z} \exp \left(  -\frac{\alpha}{\sigma} V(x) \right), \quad x \in \R,
\end{equation}
where $Z_\eps, Z > 0$ are normalization constants. Throughout this section, we will assume that the initial conditions satisfy $X_\eps(0) = X(0)$; however, we do not assume that we initialize with the invariant distribution of the processes.

The next result will be important for the proof of the asymptotic normality of the MDE; it gives weak convergence rates of the invariant densities in terms of their characteristic functions.
\begin{Lemma}  \label{lem:weak_conv_densities_langevin}
    There exists a constant $C(\alpha, \sigma, p, V, k) > 0$ such that we have the following bound for sufficiently small $\eps > 0$
    \begin{equation}   \label{eq:cf_estimate_langevin}
        \left| \E \exp(i u \xi_\eps) - \E \exp(i u \xi) \right| \leq C(\alpha, \sigma, p, V, k) \sum_{n=0}^k (1 + |u|)^n \eps^k, \quad u \in \R,
    \end{equation}
    where $\xi_\eps \sim \mu_\eps \, d \lambda^1$, $\xi \sim \mu \, d\lambda^1$.
\end{Lemma}
\begin{proof}
Corollary \ref{cor:gibbs_measure_estimates} and the estimate at the end of the proof of said corollary applied to the function $g(x) := \exp(i u x)$ with $u, x \in \R$ gives
\begin{align*}
    &\hspace{0.5cm} \left| \E \exp(i u \xi_\eps) - \E \exp(i u \xi) \right| \\
    &\leq \frac{2 \eps^k}{Z Z^- (2\pi)^k} \exp\left( \frac{1}{\sigma} \sum_{l \in \Z} |\hat{p}(l)| \right) \left\| g \exp\left(-\frac{\alpha}{\sigma} V\right) \right\|_{W^{k, 1}(\R)} \\
    &\hspace{0.5cm} + \frac{2 \eps^k}{Z^2 Z^- (2\pi)^k} \exp\left( \frac{1}{\sigma} \sum_{l \in \Z} |\hat{p}(l)| \right) \left\| \exp\left(-\frac{\alpha}{\sigma} V\right) \right\|_{W^{k, 1}(\R)} \left\| g \exp\left(-\frac{\alpha}{\sigma} V\right) \right\|_{L^1(\R)},
\end{align*}
for sufficiently small $\eps > 0$. Here, $W^{k, 1}(\R)$ is the Sobolev space of functions on $\R$ whose weak derivatives are integrable up to order $k$. Using Leibniz's rule, Hölder's inequality, and the fact that $| \partial_x^{(j)} g| \leq |u|^j$ on $\R$ for $j = 0, \ldots, k$ yields
\begin{align*}
\left\| g \exp\left(-\frac{\alpha}{\sigma} V\right) \right\|_{W^{k, 1}(\R)} 
&\leq \sum_{n=0}^k \sum_{j = 0}^n \binom{n}{k} |u|^{n-j} \left\|  \partial_x^{(j)} \exp\left(-\frac{\alpha}{\sigma} V\right) \right\|_{L^1(\R)} \\
&\leq \left\| \exp\left(-\frac{\alpha}{\sigma} V\right) \right\|_{W^{k, 1}(\R)} \sum_{n=0}^k (1 + |u|)^n,
\end{align*}
and therefore
\begin{align*}
    &\hspace{0.5cm} \left| \E \exp(i u \xi_\eps) - \E \exp(i u \xi) \right| \\
    &\leq \frac{2 \exp\left( \frac{1}{\sigma} \sum_{l \in \Z} |\hat{p}(l)| \right)}{(Z \wedge Z^2) Z^- (2\pi)^k} \left\| \exp\left(-\frac{\alpha}{\sigma} V\right) \right\|_{W^{k, 1}(\R)} \left( 1 + \left\| \exp\left(-\frac{\alpha}{\sigma} V\right) \right\|_{L^1(\R)} \right) \sum_{n=0}^k (1 + |u|)^n \eps^k \\
    &=: C(\alpha, \sigma, p, V, k) \sum_{n=0}^k (1 + |u|)^n \eps^k.
\end{align*}
\end{proof}

We now come back to the estimation problem via the MDE \eqref{eq:MDE} and fix the following $\eps$-dependent parametric model
\begin{equation}
    \mathcal{M}^{T_\eps} = \left(C([0,T_\eps];\R), \mathscr{B}^{T_\eps}, \left\{ \Pb_{\psi(\vt)}^{X_\eps} \, | \, \vartheta \in \Theta_0 \right\} \right), \quad \eps > 0,
\end{equation}
where $\mathscr{B}^{T_\eps}$ is a $\sigma$-algebra on $C([0,T_\eps];\R)$, $T_\eps \rightarrow \infty$ as $\epstozero$, and $\Theta_0 \subset (0, \infty)$ is an open, bounded interval such that its closure satisfies $\text{cl}(\Theta_0) \subset (0, \infty)$. We want to estimate the parameter $\vartheta := \alpha K$ figuring in \eqref{eq:limit_langevin} while having observations coming from \eqref{eq:multiscale_langevin}. Although we do not assume knowledge about the values of $\alpha$ and $K$, we assume that we know the value $\Bar{\sigma} := \sigma K$, e.g., from a prior estimation procedure. Such diffusion parameter estimation methods through multiscale data have been studied, for example, in \cite{AGPSZ:2021, KPK:2013, MP:2018}. The function $\psi$ is, in this case, given by $\psi(\vt) = \vt/K$. In the rest of this section, we will refer to this estimation setting as the multiscale overdamped Langevin drift parameter estimation problem, with the assumptions given at the beginning of this section. Fix again for notational convenience $\mu_\eps(\psi(\vt_0)) := \mu_\eps(\psi(\vt_0), \cdot) := \mu_\eps$ and $\mu(\vt_0) := \mu(\vt_0, \cdot) := \mu$ on $\R$ whenever $\vt_0 \in \Theta_0$ is the true parameter. 
\begin{Proposition} \label{prop:robustness_langevin}
In the multiscale overdamped Langevin drift parameter estimation problem, the family of estimators $(\hvt_{T_\eps})$ is $\eps$-robust for the true parameter $\vt_0 \in \Theta_0$.
\end{Proposition}
\begin{proof}
    Let $\vartheta_0 = \alpha_0 K \in \Theta_0$ for some $\alpha_0>0$. We will prove the validity of the conditions in Proposition \ref{prop:robustness_eps_T} where we exchange \eqref{eq:L^2-ergodic} with \eqref{eq:eps_L^2-ergodic}, recall Remark \ref{rem:remark_L^2_ergodic}. The convergence of the invariant densities follows from Lemma \ref{lem:weak_conv_densities_langevin}. Condition \eqref{eq:eps_L^2-ergodic} is a consequence of \cite[Theorem 2.10]{BK:2025} after establishing Assumptions (C) and (MET) of that same article \cite{BK:2025}, which is straightforward given the assumptions of this section. It remains to prove \eqref{eq:identifiability_cond}. To accomplish this, we assume that there exists a $\delta^*>0$ such that the identifiability condition \eqref{eq:identifiability_cond} is not satisfied. This means that we have a $\vt^* \in \Theta_0$ such that $|\vt^*-\vartheta_0|\geq\delta^*$ and $\normLebesgue{\tc[\vt^*] - \mathscr{C}_{\vt_0}}{\varphi} = 0$. This is possible by using a minimizing sequence and the fact that we assumed $\text{cl}(\Theta_0) \subset (0, \infty)$. Thus, the characteristic functions $\tc[\vt^*]$ and $\tc[\vt_0]$ coincide on $\R$ which implies
    \begin{equation*}
        \exp\left(-\frac{\vt^* - \vt_0}{\Bar{\sigma}} V(x) \right) = \frac{Z(\vt^*)}{Z(\vt_0)}, \quad x \in \R.
    \end{equation*}
    Choosing $x=0$ gives $Z(\vt^*) = Z(\vt_0)$ from which it easily follows that $\vt^*=\vt_0$, a contradiction. Hence, \eqref{eq:identifiability_cond} holds and the proof is complete.
\end{proof}

As a final result, we want to establish the asymptotic normality of $(\hvt_{T_\eps})$ in the multiscale overdamped Langevin drift parameter estimation problem. To this end, we will apply the results of the previous section and \cite[Theorem 2.19]{BK:2025}. 
\begin{Proposition} \label{prop:asy_normal_langevin}
Fix $\vartheta_0 \in \Theta_0$ as the true parameter. Let $T_\eps = \mathcal{O}(\eps^{-\gamma})$ as $\epstozero$ for some $\gamma \in (0, 2k)$. Assume that $u \mapsto |u|^{2k} \in L^1(\varphi)$ and $\normLebesgue{\partial_\vartheta \mathscr{C}_{\vartheta_0}}{\varphi}>0$. Then, in the multiscale overdamped Langevin drift parameter estimation problem, the family of estimators $(\hvt_{T_\eps})$ is asymptotically normal under the true parameter $\vartheta_0 \in \Theta_0$ as $\epstozero$, i.e., 
    \begin{equation}
    \sqrt{T_\eps} \left(\hvt_{T_\eps}(X_\eps) - \vartheta_0 \right) \weakconvunder{\psi(\vt_0)} J(\vt_0)^{-1} \mathcal{N}_1(0, \tau^2(\vt_0)), \quad \text{as } \epstozero,
    \end{equation}
    with
    \begin{equation*}
    J(\vt_0) = \normLebesgue{\partial_\vartheta \mathscr{C}_{\vartheta_0}}{\varphi}^2, \quad \tau^2(\vt_0) = 2\Bar{\sigma} \int_\R |\Phi'(x)|^2 \mu(\vt_0, x) \, dx.
    \end{equation*}
    The function $\Phi$ is given by
    \begin{equation} \label{eq:poisson_langevin}
    \Phi(x) = \int_0^x \frac{1}{\Bar{\sigma} \mu(\vt_0, y)} \int_{-\infty}^y h(z) \mu(\vt_0, z) \, dz \, dy, \quad x \in \R,
    \end{equation}
    with
    \begin{equation}    \label{eq:h_def}
        h(x) := \int_{\R} \left( \exp \left( i u x \right) - \E \exp \left( i u \xi \right) \right) \, \partial_\vt \mathscr{C}_{\vt_0}(u) \varphi(u) \, du, \quad x \in \R, \quad \xi \sim \mu(\vt_0) \, d\lambda^1.
    \end{equation}
\end{Proposition}
\begin{proof}
In the following, all appearing integrals exist due to the polynomial growth bound on $V$ and the exponential decay of $\mu$. The latter assertion is an easy consequence of \eqref{eq:ergodicity_condition} by which we have for $x > M > 0$
\begin{align}   \label{eq:potential_tail_bound}
    V(x) = V(M) + \int_M^x V'(y) \, dy \geq V(M) - \frac{b M^2}{2} + \log(x^{-a}) - \log(M^{-a}) + \frac{b}{2} x^2,
\end{align}
and similarly for $x < -M$. Furthermore, a direct calculation shows that for any $\vt \in (0, \infty)$ and $x \in \R$
\begin{align}
    \partial_\vt \, \mu(\vt, x) &= -\mu(\vt, x) \left( \frac{V(x)}{\Bar{\sigma}} + \frac{\partial_\vt Z(\vt)}{Z(\vt)} \right), \\
    \partial_\vt^2 \, \mu(\vt, x) &= \mu(\vt, x) \left[ \left( \frac{V(x)}{\Bar{\sigma}} + \frac{\partial_\vt Z(\vt)}{Z(\vt)} \right)^2 - \frac{\partial_\vt^2 Z(\vt)}{Z(\vt)} + \left( \frac{\partial_\vt Z(\vt)}{Z(\vt)} \right)^2 \right],
\end{align}
and
\begin{align}
    \partial_\vt Z(\vt) = -\frac{1}{\Bar{\sigma}} \int_\R V(y) \exp \left( -\frac{\vt}{\Bar{\sigma}} V(y) \right) \, dy, \quad
    \partial_\vt^2 Z(\vt) = \frac{1}{\Bar{\sigma}^2} \int_\R V(y)^2 \exp \left( -\frac{\vt}{\Bar{\sigma}} V(y) \right) \, dy.
\end{align}
From the preceding formulas, it is therefore easy to see that the assumptions of Proposition \ref{prop:asymptotic_normality} are satisfied. For example, it holds for $u \in \R$
\begin{equation*}
    | \partial_\vt \mathscr{C}_{\vt_0}(u) | = \left| \int_\R \exp(i u x) \, \partial_\vt \, \mu(\vt, x) \, dx \right| \leq \frac{1}{\Bar{\sigma}} \int_\R \left| V(x) \right| \mu(x) \, dx + \frac{\left| \partial_\vt Z(\vt) \right|}{Z(\vt)} < \infty.
\end{equation*}
Recall the splitting in equation \eqref{eq:asy_double_integral}
\begin{equation*}
    I(\eps) = \frac{1}{\sqrt{T_\eps}} \int_0^{T_\eps} \int_{\R} \left( \exp \left( i u X_\eps(t) \right) - \E \exp \left( i u \xi \right) \right) \, \partial_\vt \mathscr{C}_{\vt_0}(u) \varphi(u) \, du \, dt = I_1(\eps) + I_2(\eps),
\end{equation*}
with
\begin{align*}
    &I_1(\eps) = \frac{1}{\sqrt{T_\eps}} \int_0^{T_\eps} \int_{\R} \left( \exp \left( i u X_\eps(t) \right) - \E \exp \left( i u \xi_\eps \right) \right) \, \partial_\vt \mathscr{C}_{\vt_0}(u) \varphi(u) \, du \, dt, \\
    &I_2(\eps) = \sqrt{T_\eps} \int_{\R} \left( \E \exp \left( i u \xi_\eps \right) - \E \exp \left( i u \xi \right) \right) \, \partial_\vt \mathscr{C}_{\vt_0}(u) \varphi(u) \, du,
\end{align*}
where $\xi_\eps \sim \mu_\eps(\psi(\vt_0)) \, d \lambda^1$. By Lemma \ref{lem:weak_conv_densities_langevin} it follows that
\begin{align*}
    | I_2(\eps) | 
    &\leq  C(\alpha, \sigma, p, V, k) \sqrt{T_\eps} \eps^k \int_{\R} \sum_{n=0}^k (1 + |u|)^n \, | \partial_\vt \mathscr{C}_{\vt_0}(u) | \varphi(u) \, du \\
    &\leq \eps^{k-\gamma/2}  C(\alpha, \sigma, p, V, k) \int_{\R} \sum_{n=0}^k (1 + |u|)^n \, | \partial_\vt \mathscr{C}_{\vt_0}(u) | \varphi(u) \, du,
\end{align*}
so that $I_2(\eps)$ converges to zero as $\epstozero$. For the final step, we want to apply \cite[Theorem 2.19]{BK:2025} to the quantity
\begin{equation}
    \frac{1}{\sqrt{T_\eps}} \int_0^{T_\eps} h_\eps(X_\eps(t)) \, dt,
\end{equation}
with
\begin{equation}
    h_\eps(x) = \int_{\R} \left( \exp \left( i u x \right) - \E \exp \left( i u \xi_\eps \right) \right) \, \partial_\vt \mathscr{C}_{\vt_0}(u) \varphi(u) \, du, \quad x \in \R, \quad \xi_\eps \sim \mu_\eps(\psi(\vt_0)) \, d\lambda^1.
\end{equation}
First observe that Assumptions (C*), (MET), and (CLT) of \cite{BK:2025} are satisfied by our assumptions in this section. In particular, (CLT) is a consequence of inequality \eqref{eq:potential_tail_bound}. Moreover, notice that $h_\eps$ is centered with respect to the invariant density $\mu_\eps(\vt_0)$, uniformly bounded on $\R$ by $2 \| \partial_\vartheta \mathscr{C}_{\vt_0} \|_{L^1(\varphi)}$, converges uniformly on $\R$ to $h$ due to Lemma \ref{lem:weak_conv_densities_langevin}, and is continuous by the dominated convergence theorem. Finally, the condition (2.78) of \cite[Theorem 2.19]{BK:2025} is satisfied, as well; refer to Lemma \ref{lem:asymptotic_variance_convergence} for a proof. Hence, we may apply \cite[Theorem 2.19]{BK:2025} to obtain
\begin{equation}
    I_1(\eps) = \frac{1}{\sqrt{T_\eps}} \int_0^{T_\eps} h_\eps(X_\eps(t)) \, dt \weakconvunder{\psi(\vt_0)} \mathcal{N}_1(0, \tau^2(\vt_0)), \quad \epstozero,
\end{equation}
which, in turn, implies with Slutzky's Lemma and the results of Proposition \ref{prop:asymptotic_normality} and Proposition \ref{prop:robustness_langevin} the final claim.
\end{proof}
\begin{Remark}
We conclude this section with a remark regarding an extension to multidimensional processes.
The proofs of Proposition \ref{prop:robustness_langevin} and Proposition \ref{prop:asy_normal_langevin} for stochastic processes $X_\eps$ and $X$ taking values in $\R^d$ with $d>1$ and solving the equations
\begin{align}
    \label{eq:multiscale_langevin_d>1}
    &dX_\eps(t) = -\alpha \nabla V(X_\eps(t)) - \frac{1}{\eps} \nabla p\left( \frac{X_\eps(t)}{\eps} \right) dt  + \sqrt{2 \sigma} dW(t), \quad t \in [0,T], \\
    \label{eq:limit_langevin_d>1}
    &dX(t) = -\alpha K \nabla V(X(t)) dt  + \sqrt{2 \sigma K} dW(t),  \quad t \in [0,T],
\end{align}
with similar assumptions on the potential $V$ and the periodic function $p$ as in the one-dimensional case, cf. \cite{PS:2007}, pose considerable difficulties in two broader aspects. \\ \indent 
First, $T$-$\eps$- and $\eps$-$T$-robustness are actually fulfilled as long as we have the identifiability condition \eqref{eq:identifiability_cond} and second moment bounds uniform in $\eps$ as in \eqref{eq:moment_bounds_eps}. The latter may be obtained from results in \cite{BP:1978}. Ergodic theorems for these limit orders are well established in the literature, refer to \cite{DZ:1996, KMN:2012, MT:1959}. However, for the asymptotic normality result we would need $\eps$-robustness, as outlined in Remark \ref{rem:remark_L^2_ergodic}, which remains unclear thus far due to the lack of an $\eps$-dependent ergodic theorem in higher dimensions. \\ \indent
This is the first difficulty which even aggravates when taking the next step of proving an $\eps$-dependent central limit theorem for time integrals, namely
\begin{equation*}
    \frac{1}{\sqrt{T_\eps}} \int_0^{T_\eps} h_\eps(X_\eps(t)) \, dt \weakconv \mathcal{N}_d(0, \Sigma), \quad \epstozero.
\end{equation*}
Utilizing the standard method with a suitable Poisson equation, it becomes evident that we need an $\eps$-dependent ergodic theorem and sufficient bounds on the $\eps$-dependent solution of said Poisson equation, which is by no means trivial and poses the second difficulty.  \\ \indent
Last but not least, the convergence of certain asymptotic variances, given by the Dirichlet forms associated with the generators of $X_\eps$ and $X$ is closely related to an $\eps$-dependent central limit theorem for time integrals. For the considered case of Langevin dynamics, the convergence results on these asymptotic variances may be derived via results in \cite[Chapter 2]{KLO:2012}.    
\end{Remark}

\section{Numerical Experiments} \label{sec:numerics}
In this section, we will, on the one hand, substantiate the theoretical findings with numerical simulations for different example pairs of limit process and weakly perturbed version of it, and, on the other hand, empirically investigate how the MDE performs in a setting which is not fully covered by the provided theory. The simulation results have been realized in the programming language Julia, see \cite{BEKS:2017}. In particular, we used the optimization algorithms as provided by the Julia package Optim.jl, \cite{MR:2018}, for the required optimization tasks. The code can be publicly accessed through a GitHub repository \cite{B:2025}.

\subsection{Multiscale overdamped Langevin diffusion in one dimension}
Consider the multiscale overdamped Langevin drift parameter estimation problem as in Section \ref{sec:langevin} with a quadratic potential (QdP1) and quartic potential (QrP)
\begin{align}
    \label{eq:quadratic_potential_d=1}
    &V(x) = \frac12 x^2, \quad x \in \R, \tag{QdP1} \\
    \label{eq:quartic_potential}
    &V(x) = \frac14 x^4 - \frac12 x^2, \quad x \in \R. \tag{QrP}
\end{align}
When selecting the quadratic potential \eqref{eq:quadratic_potential_d=1}, the homogenized process $X$, defined through \eqref{eq:limit_langevin}, is a simple Ornstein--Uhlenbeck process and the invariant density is a Gaussian density with mean $0$ and variance $\sigma/\alpha$. Recall here that we want to estimate the true homogenized drift coefficient $\vt_0 = \alpha_0 K$ as it appears in equation \eqref{eq:limit_langevin}, while assuming that we know the homogenized diffusion coefficient $\Bar{\sigma} = \sigma K$. Hence, we replace every $\alpha$ by $\vt$, which is possible because only fractions of the form $\sigma/\alpha$ appear in the relevant terms; we then implement the optimization with respect to $\vt \in (0,\infty)$. If we take the weight function $\varphi$ as a centered Gaussian density with variance $\beta^2$, $\beta > 0$, then Example \ref{ex:general_Gaussian_dist_exmaple} gives the following general formula for $\dist$
\begin{align} \label{eq:distance_formula_gaussian_weight_1D}
\begin{aligned}
    \dist(\vt, X_\eps) = 
    &\frac{1}{T^2} \int_0^T \int_0^T k(X_\eps(t)-X_\eps(s)) \, dt \, ds \\
    &- \frac{2}{T} \int_0^T (\mu(\vt) \ast k)(X_\eps(t)) \, dt + \int_{\R} (\mu(\vt) \ast k)(x) \mu(\vt, x) \, dx\,,
\end{aligned}
\end{align}
where $k(x) = \exp(-\beta^2 x^2/2)$, $x \in \R$. Regarding the numerical implementation of the convolution terms in the time integral, we determine for a given parameter $\vartheta$ a fine spatial discretization of $\mu(\vt)$ and $k$ with a reasonable cut-off at the tails of these functions. Then we compute the convolution on this fine spatial grid with additional zero-padding using fast Fourier transforms. To get the convolution evaluated at the observation points of $X_\eps$, we apply a linear interpolation at these points. Formula \eqref{eq:distance_formula_gaussian_weight_1D} drastically simplifies in the case of a quadratic potential \eqref{eq:quadratic_potential_d=1} to
\begin{align}   \label{eq:MDE_Gaussian_1D}
\begin{aligned}
    \dist(\vt, X_\eps) = 
    &\frac{1}{T^2} \int_0^T \int_0^T \exp\left( - \frac{\beta^2}{2}(X_\eps(t)-X_\eps(s))^2 \right) \, dt \, ds \\
    &- \frac{2}{T \sqrt{1 + \beta^2 \Bar{\sigma}/ \vt }} \int_0^T \exp\left( -\frac{\beta^2 X_\eps(t)^2}{2 (1 + \beta^2 \Bar{\sigma}/ \vt)} \right) \, dt + \frac{1}{\sqrt{1 + 2 \beta^2 \Bar{\sigma}/ \vt}}.
\end{aligned}
\end{align}
We only need to consider the last two terms of the above two formulas in the implementation of the MDE since the appearing double integrals do not depend on the unknown parameter.

\begin{figure}[t]
\begin{subfigure}{0.49\textwidth}
    \includegraphics[scale=0.15, width=\textwidth]{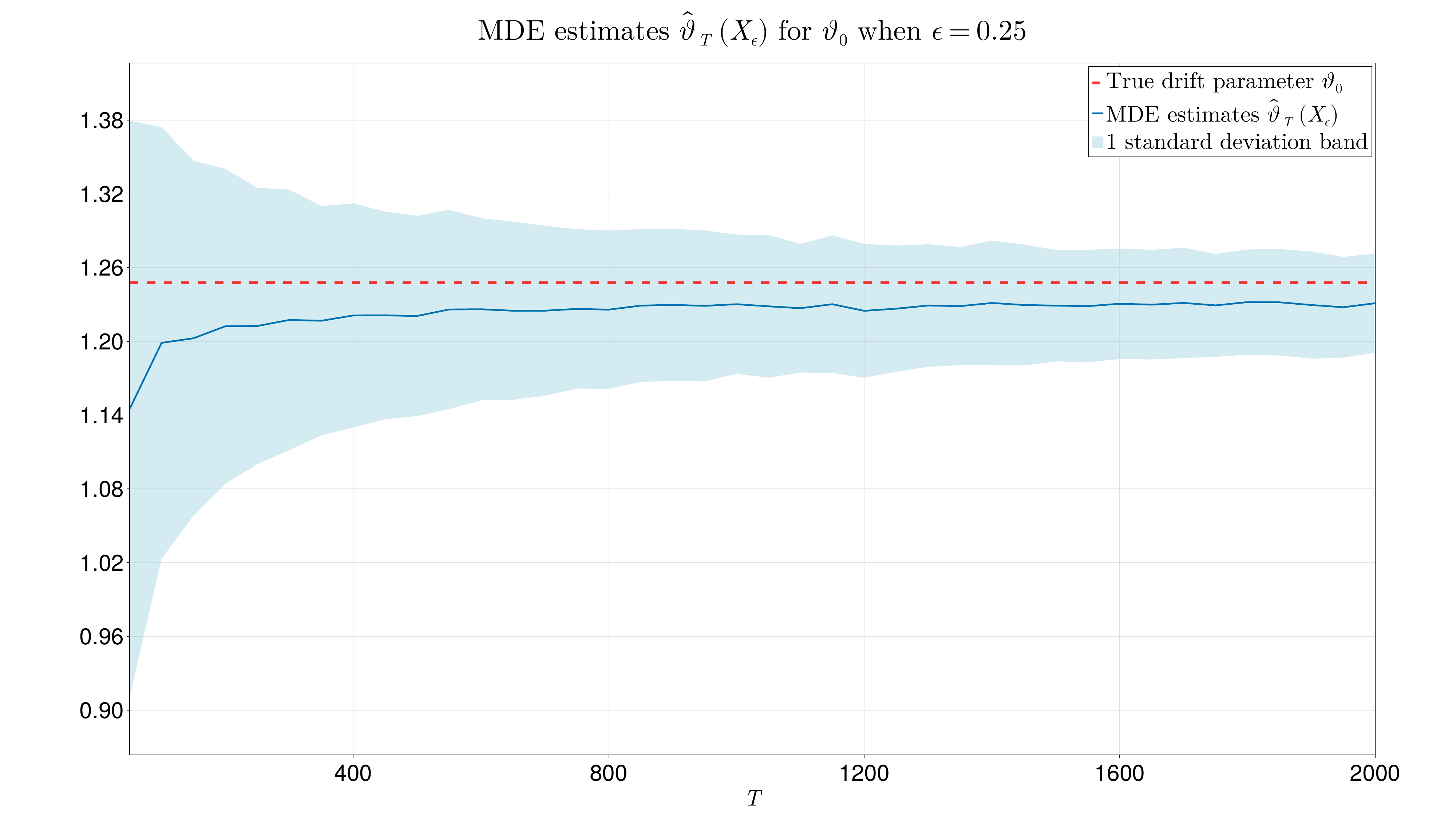}
\end{subfigure}
\begin{subfigure}{0.49\textwidth}
    \includegraphics[scale=0.15, width=\textwidth]{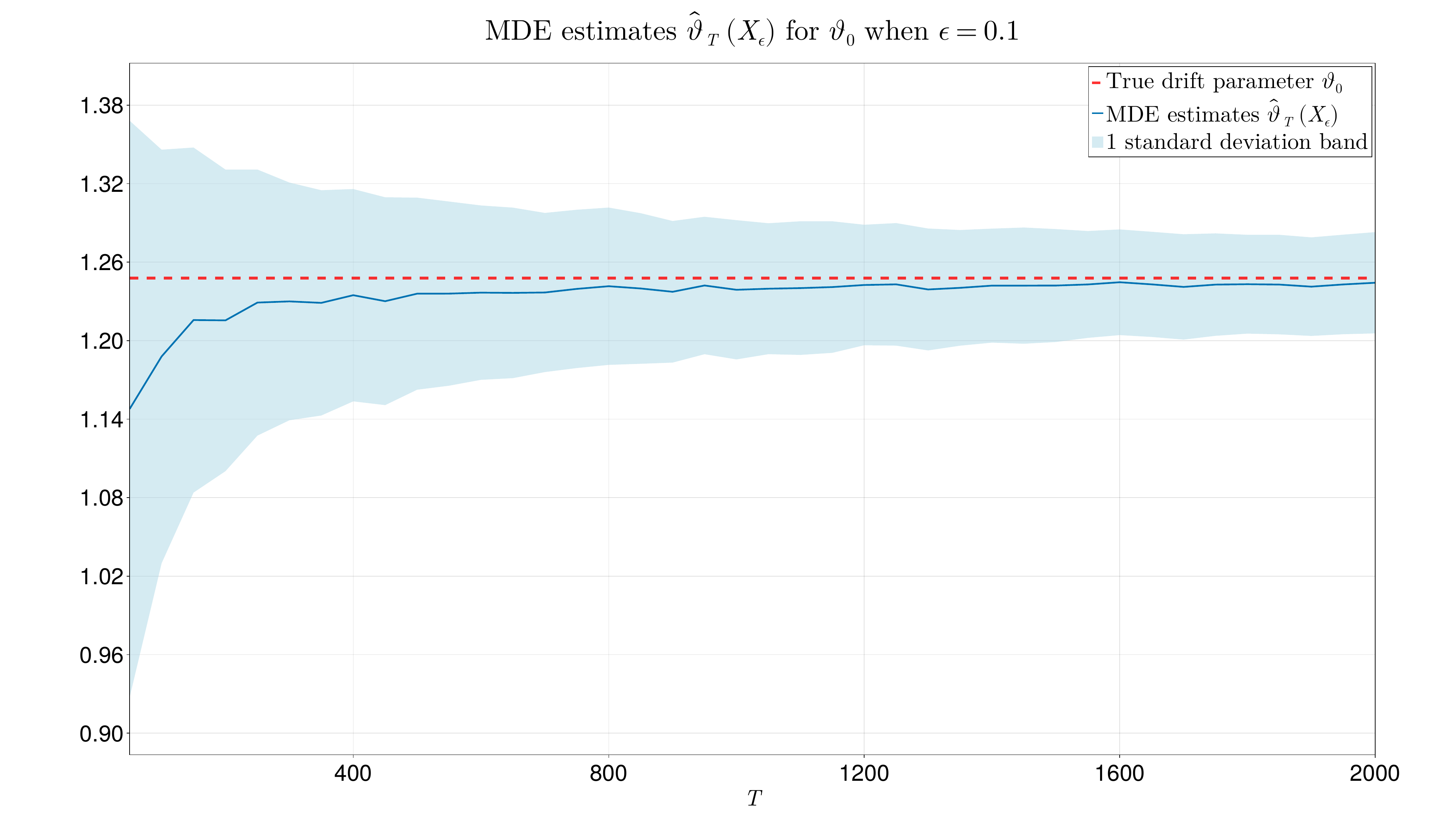}
\end{subfigure}
\caption{MDE estimates $\hvt_{T}(X_\eps)$ for $\vt_0$ when the data $X_\eps$ is a trajectory of the slow component of the multiscale overdamped Langevin diffusion \eqref{eq:multiscale_langevin} with a linear drift term corresponding to \eqref{eq:quadratic_potential_d=1}.} \label{fig:MDE_Gaussian_robustness_QP}
\end{figure}%

We generated synthetic data in the form of a time series with an ordinary Euler--Maruyama scheme using a fine time discretization $h$, plugged in the multiscale data $X_\eps$ into a discretized version of $\dist$, and minimized it with respect to $\vt$ through a (box-constrained) gradient descent algorithm. The specific parameter constellation is as follows.
\begin{itemize}
    \item Multiscale and homogenized SDE parameters:
    \begin{align*}
        \alpha_0 = 2, \quad \sigma = 1, \quad \vt_0 \approx 1.248, \quad \Bar{\sigma} \approx 0.624,    \quad X_\eps(0) = X(0) = 10.
    \end{align*}
    \item Small scale parameter and time parameters:
    \begin{align*}
        &\eqref{eq:quadratic_potential_d=1}: \eps \in \{ 0.25, 0.1 \}, \quad T \in \{ 50m \, | \, m = 1, \ldots, 40 \}, \quad h < \eps^3; \\
        &\eqref{eq:quartic_potential}: \eps = 0.1, \quad T \in \{ 100m \, | \, m = 1, \ldots, 20 \}, \quad h < \eps^3.
    \end{align*}
    \item Gradient descent algorithm parameters:
    \begin{align*}
        \beta = 1, \quad \text{initial point: } \vt_{\text{initial}} = 10, \quad  \text{box interval: } (0, \infty), \quad \text{optimizer: L-BFGS}.
    \end{align*}
\end{itemize}
Note that we make a couple of distinctions depending on the chosen drift term. This is because the optimization is numerically more involved for the quartic potential \eqref{eq:quartic_potential}. 

\begin{figure}[t]
\centering
\includegraphics[scale=0.15, width=0.8\textwidth]{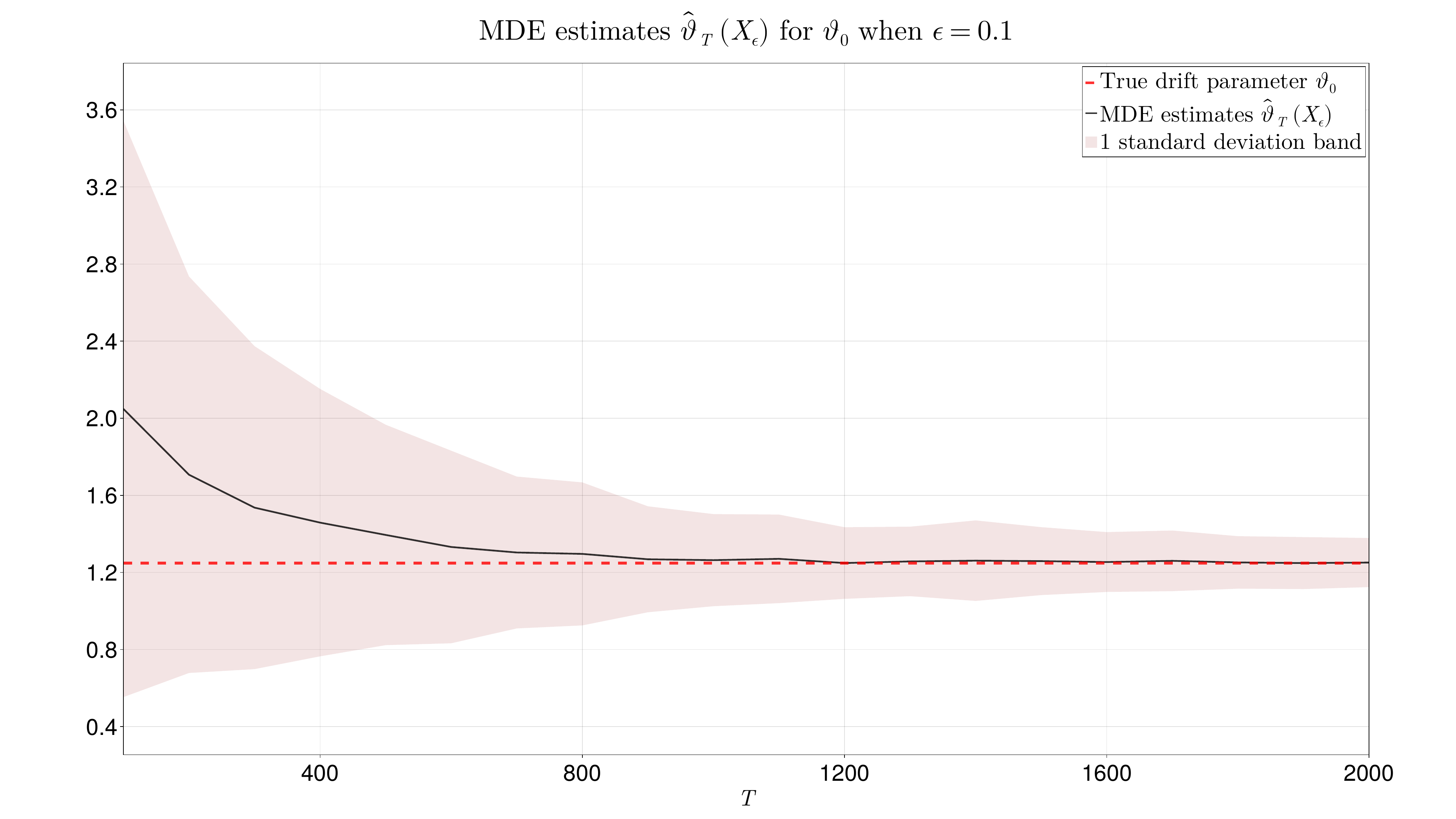}
\caption{MDE estimates $\hvt_{T}(X_\eps)$ for $\vt_0$ when the data $X_\eps$ is a trajectory of the slow component of the multiscale overdamped Langevin diffusion \eqref{eq:multiscale_langevin} with a nonlinear drift term corresponding to \eqref{eq:quartic_potential}.} \label{fig:MDE_nonlinear_robustness_BP}
\end{figure}

\begin{figure}[t]
\begin{subfigure}{0.49\textwidth}
    \includegraphics[scale=0.2, width=\textwidth]{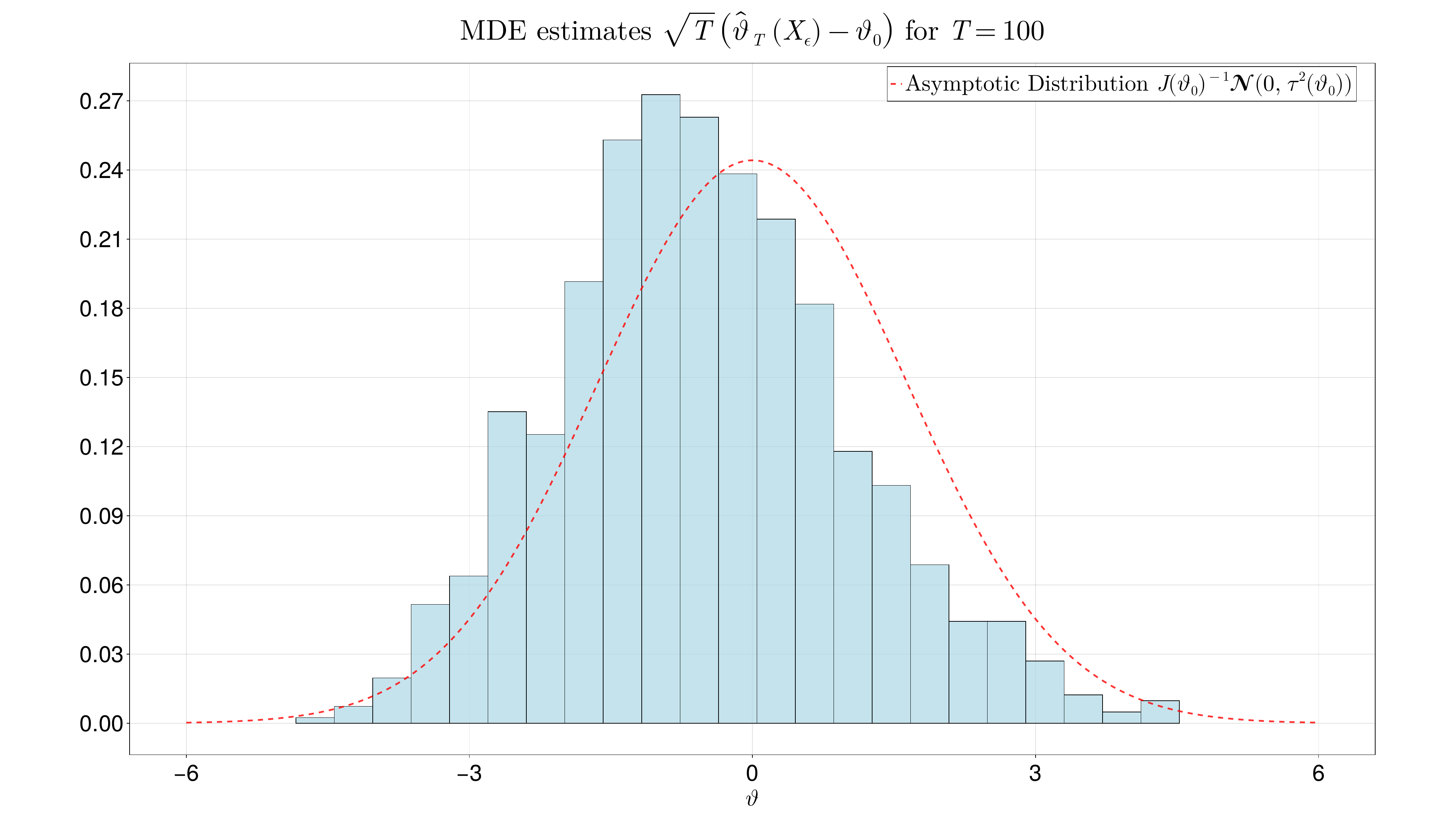}
\end{subfigure}
\begin{subfigure}{0.5\textwidth}
    \includegraphics[scale=0.2, width=\textwidth]{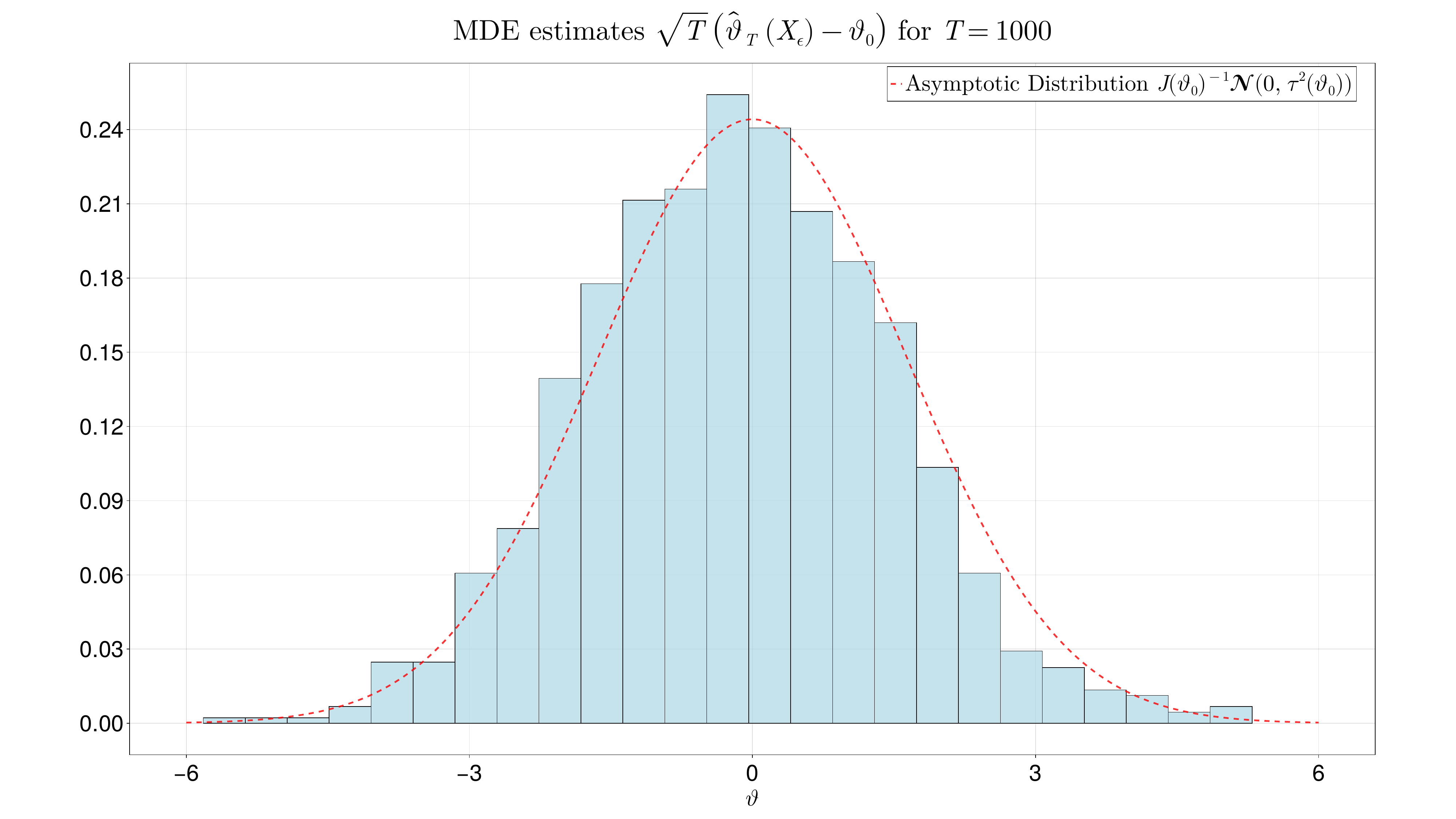}
\end{subfigure}
\caption{Normalized histogram plots of centered and scaled MDE estimates $\sqrt{T} ( \hvt_{T}(X_\eps) - \vt_0 )$ for $\vt_0$ when the data $X_\eps$ is a trajectory of the slow component of the multiscale overdamped Langevin diffusion \eqref{eq:multiscale_langevin} with a linear drift term corresponding to \eqref{eq:quadratic_potential_d=1}. The histogram bin number has been determined according to the Freedman--Diaconis rule.} \label{fig:MDE_Gaussian_normality_QP}
\end{figure}

Figure \ref{fig:MDE_Gaussian_robustness_QP} and \ref{fig:MDE_nonlinear_robustness_BP} show estimation results regarding the robustness of the MDE in the case \eqref{eq:quadratic_potential_d=1} and \eqref{eq:quartic_potential}. Here, the estimated values are, in fact, averages of the same experiment repeated for 1000 times. We can observe that the MDE $\hvt_T(X_\eps)$ succeeds in estimating the true parameter $\vt_0$ for decreasing $\eps$ and increasing $T$. This is in agreement with the established theoretical results of Section \ref{sec:langevin}. \\ \indent
Moreover, we briefly want to draw attention to numerical results regarding the asymptotic normality of the MDE. We consider the case \eqref{eq:quadratic_potential_d=1} with the same parameter configuration as before, but this time only for $\eps=0.1$ and $T \in \{100, 1000\}$. The two plots in Figure \ref{fig:MDE_Gaussian_normality_QP} show normalized histograms of 1000 estimates, together with the (approximated) asymptotic Gaussian density of Proposition \ref{prop:asy_normal_langevin}.
A short calculation reveals that
\begin{equation}
    J(\vt_0) = \frac{3}{4 \beta} \left( \frac{\Bar{\sigma}}{\vt_0^2} \right)^2 \sigma_2^5, \quad h(z) = \frac{\Bar{\sigma}}{2 \vt_0^2 \beta} \left[ \sigma_1^3 (1 - \sigma_1^2 z^2) \exp \left( -\frac{\sigma_1^2 z^2}{2} \right) - \sigma_2^3 \right], \quad z \in \R,
\end{equation}
with
\begin{equation}
    \sigma_1^2 := \frac{\beta^2 \vt_0}{\vt_0 + \Bar{\sigma}\beta^2 }, \quad \sigma_2^2 := \frac{\beta^2 \vt_0}{\vt_0 + 2 \Bar{\sigma}\beta^2 }.
\end{equation}
These formulas facilitate the numerical approximation of the asymptotic variance. In this particular case, we calculated an approximate value of $\tau^2(\vt_0)/J(\vt_0)^2 \approx 2.670$. As we can see from the figures, the results are in relatively good agreement with the theoretical results.

\subsection{Multiscale overdamped Langevin diffusion in two dimensions}
Consider the equations \eqref{eq:multiscale_langevin_d>1} and \eqref{eq:limit_langevin_d>1} for $d=2$ with 
\begin{equation}    \label{eq:quadratic_potential_d=2}
    V(x) = \frac12 x^\top M x, \quad 
    p(x) := p_1(x_1) + p_2(x_2) := \sin(x_1) + \frac12 \sin(x_2) \tag{QrP2}
\end{equation}
where $x:=(x_1, x_2)^\top \in \R^2$ and $M \in \R^{2 \times 2}$ is a positive definite matrix. In this case, the equations read as follows for $X_\eps := (X^{(1)}_\eps, X^{(2)}_\eps)^\top$
\begin{align}
    \label{eq:multiscale_langevin_2d}
    &dX_\eps(t) = -\alpha M X_\eps(t) - \frac{1}{\eps} 
    \begin{pmatrix}
        \cos\left( X^{(1)}_\eps(t)/\eps \right) \\[0.1cm]
        \frac12 \cos\left( X^{(2)}_\eps(t)/\eps \right)
    \end{pmatrix} dt  + \sqrt{2 \sigma} dW(t), \\[0.5cm]
    \label{eq:limit_langevin_2d}
    &dX(t) = -\alpha K M X(t) dt  + \sqrt{2 \sigma K} dW(t),
\end{align}
and the corrective constant $K$ equals 
\begin{equation}
    K :=
    \begin{pmatrix}
        k_1 &0 \\ 0 &k_2
    \end{pmatrix}
    :=
    \begin{pmatrix}
        (Z_1^+ Z_1^-)^{-1} &0 \\ 0 & (Z_2^+ Z_2^-)^{-1}
    \end{pmatrix}
    , \quad Z_i^{\pm} := \frac{1}{2 \pi} \int_0^{2 \pi} \exp \left( \pm \frac{p_i(y)}{\sigma} \right) \, dy, \quad i = 1,2.
\end{equation}
We set $\alpha = 1$ and formulate our goal as the estimation of the positive definite matrix $\vt_0 := KM_0$ for some positive definite matrix $M_0 \in \R^{2 \times 2}$ while assuming that we know the matrix $\Sigma := \sigma K$. The invariant density $\mu(\vt_0)$ is the density of a centered 2-dimensional normal distribution with covariance matrix
\begin{equation}
    \sigma M_0^{-1} = (KM_0)^{-1} \sigma K = \vt_0^{-1} \Sigma. 
\end{equation}
If we choose the weight $\varphi$ as in Example \ref{ex:general_Gaussian_dist_exmaple}, then we can utilize the simplified formula \eqref{eq:distance_formula_normal_distr} for the optimization task. The optimization was performed using an interior-point Newton algorithm with suitable nonlinear constraints ensuring that the objective functional of the optimization, namely $\dist$, is well-defined. To be more precise, we optimized with respect to a matrix $A \in \R^{2 \times 2}$ that satisfies the following constraints
\begin{align}   \label{eq:constraints}
    \begin{aligned}
        &c_1(A) := A_{11} > 0, \\
        &c_2(A) := A_{11} A_{22} - A_{21} A_{12} > 0, \\
        &c_3(A) := \frac{A_{12}}{\Sigma_{11}} - \frac{A_{21}}{\Sigma_{22}} = 0.
    \end{aligned}
\end{align}
These constraints are derived from the requirements that $A$ and $A^{-1} \Sigma$ must be positive definite matrices. The exact parameter constellation that we used is as follows.
\begin{itemize}
    \item Multiscale and homogenized SDE parameters:
    \begin{align*}
        &M_0 = 
        \begin{pmatrix}
            4 &2 \\
            2 &3
        \end{pmatrix},
        \quad \sigma = 1.5, 
        \quad \vt_0 
        \approx 
        \begin{pmatrix}
            3.222 &1.611 \\
            1.893 &2.839
        \end{pmatrix}, 
        \quad \Sigma 
        \approx 
        \begin{pmatrix}
            1.208 &0 \\
            0     &1.419
        \end{pmatrix}, \\
        &X_\eps(0) = X(0) = 
        \begin{pmatrix}
            10 \\
            10
        \end{pmatrix}.
    \end{align*}
    \item Small scale parameter and time parameters:
    \begin{align*}
        \eps = 0.1, \quad T \in \{ 50k \, | \, k = 1, \ldots, 20 \}, \quad h = 10^{-3}.
    \end{align*}
    \item Interior-point Newton algorithm parameters:
    \begin{align*}
        \beta = 1, \quad \text{initial point: } \vt_{\text{initial}} =
        \begin{pmatrix}
            3 &\Sigma_{11}/2 \\
            \Sigma_{22}/2 &6
        \end{pmatrix},
        \quad \text{optimizer: IPNewton}.
    \end{align*}
\end{itemize}

\begin{figure}[t]
\centering
\includegraphics[width=\textwidth]{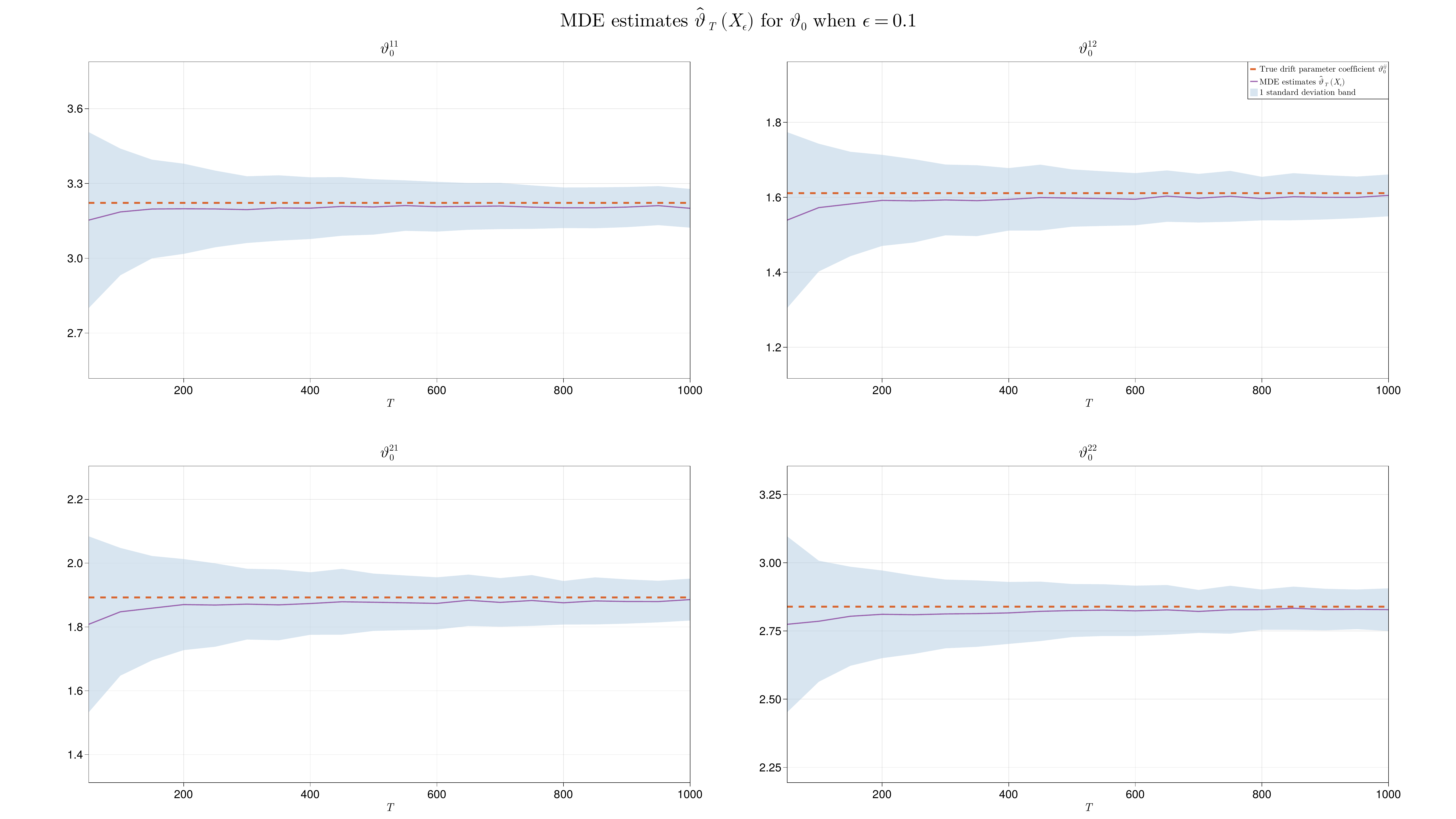}
\caption{MDE estimates $\hvt_{T}(X_\eps)$ for $\vt_0$ when the data $X_\eps$ is a trajectory of the slow component of the multiscale overdamped Langevin diffusion \eqref{eq:multiscale_langevin_2d} with a linear drift term corresponding to \eqref{eq:quadratic_potential_d=2}.} \label{fig:MDE_Gaussian_robustness_2D}
\end{figure}%

Figure \ref{fig:MDE_Gaussian_robustness_2D} shows the estimation results of the same experiment repeated 500 times. We can once again see that the MDE is capable of estimating the true parameter, which validates the robustness results for a multidimensional example.
\subsection{Fast Chaotic Noise}
Consider now the following system of ordinary differential equations (ODEs) with a slow component $X_\eps$ and a fast component $Y_\eps := (Y_\eps^{(1)}, Y_\eps^{(2)}, Y_\eps^{(3)} )$
\begin{align}   \label{eq:fast_chaotic_noise_system}
    \begin{aligned}
        \frac{dX_\eps}{dt} &= AX - BX^3 + \frac{\lambda}{\eps} Y_\eps^{(2)}, \\
        \frac{dY_\eps^{(1)}}{dt} &= \frac{10}{\eps^2} \left( Y_\eps^{(2)} - Y_\eps^{(1)} \right), \\
        \frac{dY_\eps^{(2)}}{dt} &= \frac{1}{\eps^2} \left( 28 Y_\eps^{(1)} - Y_\eps^{(2)} - Y_\eps^{(1)}Y_\eps^{(3)} \right), \\
        \frac{dY_\eps^{(3)}}{dt} &= \frac{1}{\eps^2} \left( Y_\eps^{(1)}Y_\eps^{(2)} - \frac83 Y_\eps^{(3)} \right),
    \end{aligned}
\end{align}
where $Y_\eps$ solves the Lorenz equation and $A, B \in \R$, and $\lambda > 0$. By eliminating the fast component $Y_\eps$, the authors of \cite{GKS:2004} showed that the slow dynamics $X_\eps$ can be approximated by the solution of an SDE given by
\begin{equation}  \label{eq:fast_chaotic_noise_limit}
    dX = (AX - BX^3)dt + \sqrt{\Bar{\sigma}} dW_t. 
\end{equation}
Here, the diffusion coefficient $\Bar{\sigma}$ can be expressed with the Green--Kubo formula
\begin{equation}    \label{eq:green-kubo_diffusion}
    \Bar{\sigma} = \frac{\lambda^2}{2} \int_0^\infty \lim_{T \rightarrow \infty} \frac1T \int_0^T Y_{\eps=1}^{(2)}(s) Y_{\eps=1}^{(2)}(s+t) \, ds \, dt.
\end{equation}
A rigorous justification of this approximation can be found in \cite{MS:2011}. Considering the computational cost of approximating this diffusion coefficient numerically, we will use the MDE for the estimation and plug in observations coming from the slow component $X_\eps$ of the system \eqref{eq:fast_chaotic_noise_system}. This is possible because, when keeping the formulas \eqref{eq:distance_formula_gaussian_weight_1D} and \eqref{eq:MDE_Gaussian_1D} in mind, it is immediate that there is no need to change anything about the methodology. We merely have to optimize these distance formulas with respect to the parameter $\Bar{\sigma}$ now. However, we again have to assume that the true drift coefficients $A$ and $B$ are given or obtained from a prior estimation procedure.

Due to the stiffness of the ODE system \eqref{eq:fast_chaotic_noise_system} and the deterministic nature of it, a simple fourth order Runge--Kutta scheme is sufficient to generate a trajectory of the slow component. The considered parameter constellation in this example is as follows.
\begin{itemize}
    \item ODE parameters:
    \begin{align}
        &A = -1, \quad B = 0, \quad \lambda = \frac{2}{45}, \quad \left( X_\eps(0), Y_\eps(0) \right) = (1, 1, 1, 1); \\
        &A = B = 1, \quad \lambda = \frac{2}{45}, \quad \left( X_\eps(0), Y_\eps(0) \right) = (1, 1, 1, 1).
    \end{align}
    \item Small scale parameter and time parameters:
    \begin{align*}
        &\eps \in \{10^{-1}, 10^{-3/2}\}, \quad T \in \{ 100, 500, 1000 \}, \quad h = 10^{-3}.
    \end{align*}
    \item Gradient descent algorithm parameters:
    \begin{align*}
        \beta = 1, \quad \text{initial point: } \vt_{\text{initial}} = 0.8, \quad  \text{box interval: } (0, \infty), \quad \text{optimizer: L-BFGS}.
    \end{align*}
\end{itemize}
\begin{table}[t]
\centering
\begin{tabular}{>{\hspace{2mm}}c<{\hspace{2mm}}|r|rrr>{\hspace{2mm}}c<{\hspace{2mm}}|r|rrrr}
    \toprule[0.3mm]
    \midrule
                        & \backslashbox{$\eps$}{$T$}                    
                        & 100   & 500   & 1000  &                          & \backslashbox{$\eps$}{$T$}                                                                            & 100   & 500   & 1000      \\ \midrule
$A=-1$, & $10^{-1}$     & 0.131 & 0.115 & 0.118 &   $A=1$,  & $10^{-1}$    & 0.069 & 0.104 & 0.105     \\
$B=0$   & $10^{-3/2}$   & 0.123 & 0.118 & 0.109 &   $B=1$   & $10^{-3/2}$  & 0.100 & 0.127 & 0.118     \\ 
    \midrule
    \bottomrule[0.3mm]
\end{tabular}
\caption{MDE estimates $\hvt_{T}(X_\eps)$ for $\Bar{\sigma}_0$ when the data $X_\eps$ is a trajectory of the slow component of the system \eqref{eq:fast_chaotic_noise_system} under different parameter configurations.}   \label{table:fcn}
\end{table}%

Table \ref{table:fcn} shows the estimated values for the true diffusion coefficient $\vt_0 := \Bar{\sigma}_0$ under the specified parameter configurations. Since there is no analytic expression for $\Bar{\sigma}_0$, we may compare the obtained estimates with the results given in \cite{GKS:2004}. There, they adopted a setting with $A=B=1$ and $\eps = 10^{-3/2}$, giving an estimate of $0.126 \pm 0.003$ via Gaussian moment approximations based on a modified Euler–-Maruyama discretization of the limit equation \eqref{eq:fast_chaotic_noise_limit}, and an estimate of $0.13 \pm 0.01$ based on the heterogeneous multiscale method, see \cite{WLV:2005} and the references therein for more details on this method, with a discretization of the Green–-Kubo formula for $\Bar{\sigma}_0$. When $T \in \{500, 1000\}$, we can observe that we obtain similar values for the diffusion coefficient $\Bar{\sigma}_0$, which, at least empirically, solidifies the robustness of the MDE for an example that is not covered by the theoretical results.

\section{Conclusions}       \label{sec:conclusion}
In this work, we presented a general method for estimating parameters in models that are known up to the sought-after parameters. The approach leads to a plug-in minimum distance estimator that minimizes a certain weighted $L^2$-norm. The gist of the work lies in the permission of a certain misspecification between data and model, that is, we do not assume that the data, manifesting itself in form of one long trajectory, comes from the true model, but rather from a model that is, in a sense, a weak perturbation of the true model, controlled by a small scale parameter $\eps$. Throughout the work, we assumed that the law of the weakly perturbed model converges to the law of the true model as $\epstozero$. We then studied and proved the robustness of the estimator as the small scale parameter goes to zero and the final trajectory time goes to infinity in different limit orders. We also studied the asymptotic normality of the estimator, which we were able to prove at least for a class of multiscale overdamped Langevin diffusions in one dimension. Some numerical simulation results have been presented in order to substantiate the theoretical results.

Our theoretical and numerical results also reveal some limitations of the methodology, which, of course, sparks motivation for possible further research. From a theoretical point of view, immediate open questions are extensions and generalizations of the proofs, especially the proof of the asymptotic normality of the estimator, to other classes of stochastic processes, e.g., Lévy processes. Even though most of the proof ideas are quite general and, in parts, even standard, getting the right estimates for a rigorous justification poses formidable challenges because terms may very well be unbounded with respect to the small scale parameter $\eps$ or the time horizon $T$, which, as a consequence, necessitates a more sophisticated analysis and tracking of the relevant terms with respect to $\eps$ and $T$. Concerning the numerical implementation of the estimation method, one of the most striking obstacles of the MDE is the inherent optimization problem that one needs to solve to get a plausible estimate. This can be extremely difficult as soon as we exit the setting of Gaussian densities. Hence, it is certainly crucial to have more elaborate and sophisticated numerical implementations, especially global optimization routines, in order to expand the practicability of the approach. Another limitation concerns the estimation setting itself, namely, we specifically require knowledge about the true diffusion coefficient in order to avoid identifiability issues. We want to remind the reader that this is, as a matter of fact, in the setting of estimation through multiscale data not a trivial problem that merely reduces to the estimation via the classical quadratic variation estimator, see, for example, \cite[Section 2.3]{K:2014}. This seems to be a larger gap in the existing literature on diffusion estimation via multiscale observations, although there have been successful attempts at resolving this issue as in \cite{KPK:2013, MP:2018, AGPSZ:2021}. A final interesting question would be the extension of the proposed methodology to a discrete setting with only a limited number of data points, which is arguably a more realistic setting from a practitioner's point of view. 

\appendix
\section{Technical results}    \label{app:A}

In this section, we want to derive an estimate for an integral that is related to certain oscillatory integrals and which produces quantitative weak convergence rates for a special instance of invariant measures. We refer the reader to the standard text \cite[Chapter 3]{G:2008} for a detailed presentation of Fourier analysis on the $d$-dimensional torus $\T$ as we make heavy use of this machinery in the following.

Let $k \in \N$. Consider the functions $f \in C^k(\R^d; \R)$ with sufficient decrease, i.e., for any multiindex $\beta \in \N_0^d$ with $| \beta | \leq k$ there exists a constant $C_\beta > 0$ such that
\begin{equation}
    | (\partial^\beta f) (x) | \leq \frac{C_\beta}{1 + | x |_2^{d+1}}, \quad x \in \R^d.
\end{equation}
We denote the set of these functions by $S^k(\R^d)$. It is clear that $S^k(\R^d) \subset W^{k, 1}(\R^d)$, the Sobolev space of functions whose weak derivatives are integrable up to order $k$, and that $x \mapsto \exp(-|x|_2^2) \in S^k(\R^d)$ for any $k \in \N$. 
\begin{Definition} \label{def:potential}
Let $k \in \N$. We call a function $V \colon \R^d \to \R$ a potential whenever it satisfies
\begin{enumerate}[label=\roman*)]
    \item $V \in C^k(\R^d; \R)$.
    \item $V$ has a global lower bound and grows at most polynomially, i.e., there exists an $L \in \R$ and a polynomial $P \colon \R^d \to \R$ of some degree such that
    \begin{equation*}
        V(x) \geq L, \quad \left| V(x) \right| \leq P(x), \quad x \in \R^d.
    \end{equation*}
    \item There exist positive numbers $a, b > 0$ such that
    \begin{equation*}
        - \nabla V(x)^\top x \leq a - b |x|_2^2, \quad x \in \R^d.
    \end{equation*}
\end{enumerate}
\end{Definition}
\begin{Remark}  \label{rem:V_potential_estimate}
Let $k \in \N$. If $V$ is a potential, then $x \mapsto \exp(-V(x)) \in S^k(\R^d)$.
\end{Remark}
\begin{proof}
Using the fundamental theorem of calculus for line integrals and Assumptions \ref{def:potential} ii) and iii) gives for $x \in \R^d$
\begin{align*}
    V(x) 
    &= V(0) + \int_0^1 \nabla V(tx)^\top x \, dt = V(0) + \int_0^{1/2} \nabla V(tx)^\top x \, dt + \int_{1/2}^1 \nabla V(tx)^\top (tx) \, \frac{dt}{t} \\
    &\geq V(x/2) + \int_{1/2}^1 \left( b t^2 |x|_2^2 - a \right) \frac{dt}{t} \geq L - a \log(2) + \frac{3b}{8} |x|_2^2,
\end{align*}
which, in turn, yields
\begin{equation}    \label{eq:potential_exp_estimate}
    \exp(-V(x)) \leq 2^a \exp(-L) \exp\left( -\frac{3b}{8} |x|_2^2 \right), \quad x \in \R^d.
\end{equation}
Arbitrary derivatives of $x \mapsto \exp(-V(x))$ up to order $k$ have similar exponential bounds so that the claim follows.
\end{proof}

If $\lambda \in \R^d \backslash \{0\}$, then successive integration by parts gives the formula
\begin{equation}    \label{eq:fourier_transform_diff}
    \int_{\R^d} (\partial^\beta f)(x) \exp\left( 2 \pi i \lambda^\top x \right) dx = (-1)^{|\beta|} (2 \pi i \lambda)^\beta \int_{\R^d} f(x) \exp\left( 2 \pi i \lambda^\top x \right) dx,
\end{equation}
for any multiindex $\beta \in \N_0^d$ and $f \in S^k(\R^d)$. This is obviously just the differentiation rule for the Fourier transform on $\R^d$ and will be useful for what follows. We now come to the previously announced integral estimate.
\begin{Proposition} \label{prop:oscillatory_integral_estimate}
Let $p \colon \T \to \R$ be a continuous $1$-periodic function such that its Fourier series is absolutely summable:
\begin{equation}
    \sum_{l \in \Z^d} |\hat{p}(l)| < \infty, \quad \hat{p}(l) := \int_{\T} p(y) \exp\left( 2 \pi i l^\top y \right) dy, \quad l \in \Z^d.
\end{equation}
Then for all $\eps > 0$ and $f \in S^k(\R^d)$
\begin{equation} \label{eq:oscillatory_integral_estimate}
    \left| \int_{\R^d} f(x) \left[ \exp \left( p\left( x/\eps \right) \right) - \int_{\T} \exp \left( p(y) \right) dy  \right] dx \right| \leq \frac{\eps^k}{(2\pi)^k} \exp\left( \sum_{l \in \Z^d} |\hat{p}(l)| \right) \| f \|_{W^{k, 1}(\R^d)}.
\end{equation}
\end{Proposition}
\begin{proof}
For the beginning we consider the case of a general trigonometric polynomial on the torus given by 
\begin{equation*}
    T_N(x) := \sum_{\substack{l \in \Z^d \\[0.1cm] |l| \leq N}} a(l) \exp(2 \pi i l^\top x), \quad x \in \T, \quad N \in \N\,,
\end{equation*}
with coefficients $a(l) \in \C$. We will occasionally suppress writing $l \in \Z^d$ and likewise notation in the appearing sums whenever it does not create confusion. Given the set of coefficient indices $A := \{ l \in \Z^d \, | \, |l| \leq N \}$ there exists a bijective map
\begin{equation*}
    I \colon A \to \{1, \ldots, M\}; \quad l \mapsto I(l),
\end{equation*}
where $M := |A|$ is the cardinality of $A$. Using a series expansion and the multinomial theorem we can write
\begin{align*}
    \int_{\T} \exp \left( T_N(y) \right) dy &= \sum_{n=0}^\infty \frac{1}{n!} \int_{\T} \left( T_N(y) \right)^n dy \\
    &= \sum_{n=0}^\infty \frac{1}{n!} \int_{\T} \left( \sum_{m=1}^M a(I^{-1}(m)) \exp(2 \pi i I^{-1}(m)^\top y) \right)^n dy \\
    &= \sum_{n=0}^\infty \frac{1}{n!} \sum_{\substack{\alpha \in \N_0^M \\[0.1cm] |\alpha| = n}} \binom{n}{\alpha} \int_{\T} \prod_{m=1}^M \left[ a(I^{-1}(m)) \exp(2 \pi i I^{-1}(m)^\top y) \right]^{\alpha_m} dy \\
    &= \sum_{n=0}^\infty \frac{1}{n!} \sum_{\substack{|\alpha| = n}} \binom{n}{\alpha} \prod_{m=1}^M a(I^{-1}(m))^{\alpha_m} \int_{\T} \left[ \exp(2 \pi i I^{-1}(m)^\top y) \right]^{\alpha_m} dy \\
    &= \sum_{n=0}^\infty \frac{1}{n!} \sum_{\substack{|\alpha| = n}} \binom{n}{\alpha} \prod_{m=1}^M a(I^{-1}(m))^{\alpha_m} \int_{\T} \exp\left( 2 \pi i y^\top \sum_{m=1}^M I^{-1}(m)\alpha_m \right) dy.
\end{align*}
We set $S^M_\alpha := \sum_{m=1}^M I^{-1}(m)\alpha_m$ and note that $S^M_\alpha \in \Z^d$ and by virtue of the periodicity of the complex exponential
\begin{equation*}
    \int_{\T} \exp\left( 2 \pi i y^\top S^M_\alpha \right) dy = \mathds{1}_{\{S^M_\alpha=0\}},
\end{equation*}
so that the above calculation simplifies to
\begin{equation}    \label{eq:integral_formula}
    \int_{\T} \exp \left( T_N(y) \right) dy = \sum_{n=0}^\infty \frac{1}{n!} \sum_{\substack{|\alpha| = n \\[0.1cm] S^M_\alpha = 0}} \binom{n}{\alpha} \prod_{m=1}^M a(I^{-1}(m))^{\alpha_m}.
\end{equation}
With similar steps as before we also obtain for any $\eps > 0$ and $x \in \R^d$
\begin{equation*}
    \exp \left( T_N(x/\eps) \right) = \sum_{n=0}^\infty \frac{1}{n!} \sum_{\substack{\alpha \in \N_0^M \\[0.1cm] |\alpha| = n}} \binom{n}{\alpha} \prod_{m=1}^M a(I^{-1}(m))^{\alpha_m} \exp\left( 2 \pi i y^\top (S^M_\alpha/\eps) \right).
\end{equation*}
Observe that
\begin{align*}
    \exp \left( T_N(x/\eps) \right) 
    &= \sum_{n=0}^\infty \frac{1}{n!} \sum_{\substack{|\alpha| = n \\[0.1cm] S^M_\alpha = 0}} \binom{n}{\alpha} \prod_{m=1}^M a(I^{-1}(m))^{\alpha_m} \\
    &\hspace{0.25cm} + \sum_{n=0}^\infty \frac{1}{n!} \sum_{\substack{|\alpha| = n \\[0.1cm] S^M_\alpha \neq 0}} \binom{n}{\alpha} \prod_{m=1}^M a(I^{-1}(m))^{\alpha_m} \exp\left( 2 \pi i x^\top (S^M_\alpha/\eps) \right) \\
    &= \int_{\T} \exp \left( T_N(y) \right) dy + \sum_{n=0}^\infty \frac{1}{n!} \sum_{\substack{|\alpha| = n \\[0.1cm] S^M_\alpha \neq 0}} \binom{n}{\alpha} \prod_{m=1}^M a(I^{-1}(m))^{\alpha_m} \exp\left( 2 \pi i x^\top (S^M_\alpha/\eps) \right).
\end{align*}
For the next step we note that for any multiindex $\alpha \in \N_0^M$ with $|\alpha| = n$ and $S^M_\alpha \neq 0$ there exist indices $j_1, \ldots, j_m \in \{1, \ldots, d\}$, $m \leq d$, such that $(S^M_\alpha)_{j_1}, \ldots, (S^M_\alpha)_{j_m} \neq 0$. Hence, using formula \eqref{eq:fourier_transform_diff} we obtain the following estimate for any multiindex $\beta \in \N_0^d$ with $|\beta| = \sum_{j=1}^d \beta_j = \sum_{i=1}^m \beta_{j_i} = k$
\begin{equation}
    \left| \int_{\R^d} f(x) \exp\left( 2 \pi i x^\top (S^M_\alpha/\eps) \right) dx \right|
    \leq \frac{\eps^k \| f \|_{W^{k, 1}(\R^d)}}{(2\pi)^k |(S^M_\alpha)^\beta|} \leq \frac{\eps^k \| f \|_{W^{k, 1}(\R^d)}}{(2\pi)^k},
\end{equation}
where we used the fact that $|(S^M_\alpha)^\beta| > 1$ in the last inequality.
Using this bound and the preceding calculations we get
\begin{align*}
    &\hspace{0.5cm} \left| \int_{\R^d} f(x) \left[ \exp \left( T_N\left( x/\eps \right) \right) - \int_{\T} \exp \left( T_N(y) \right) dy  \right] dx \right| \\ 
    &\leq \sum_{n=0}^\infty \frac{1}{n!} \sum_{\substack{|\alpha| = n \\[0.1cm] S^M_\alpha \neq 0}} \binom{n}{\alpha} \prod_{m=1}^M \left| a(I^{-1}(m)) \right|^{\alpha_m} \left| \int_{\R^d} f(x) \exp\left( 2 \pi i y^\top (S^M_\alpha/\eps) \right) dy \right| \\
    &\leq \frac{\eps^k \| f \|_{W^{k, 1}(\R^d)}}{(2\pi)^k} \sum_{n=0}^\infty \frac{1}{n!} \sum_{\substack{|\alpha| = n \\[0.1cm] S^M_\alpha \neq 0}} \binom{n}{\alpha} \prod_{m=1}^M \left| a(I^{-1}(m)) \right|^{\alpha_m}.
\end{align*}
By another application of the multinomial theorem it holds
\begin{align*}
    \sum_{\substack{|\alpha| = n \\[0.1cm] S^M_\alpha \neq 0}} \binom{n}{\alpha} \prod_{m=1}^M \left| a(I^{-1}(m)) \right|^{\alpha_m}
    &\leq \sum_{\substack{|\alpha| = n}} \binom{n}{\alpha} \prod_{m=1}^M \left| a(I^{-1}(m)) \right|^{\alpha_m} \\
    &= \left( \sum_{m=1}^M  \left| a(I^{-1}(m)) \right| \right)^n
    = \left( \sum_{\substack{|l| \leq N}}  \left| a(l) \right| \right)^n.
\end{align*}
Hence, we obtain the following estimate for a general trigonometric polynomial
\begin{equation}    \label{eq:estimate_trig_poly}
    \left| \int_{\R^d} f(x) \left[ \exp \left( T_N\left( x/\eps \right) \right) - \int_{\T} \exp \left( T_N(y) \right) dy  \right] dx \right| \leq \frac{\eps^k }{(2\pi)^k} \exp \left( \sum_{\substack{|l| \leq N}}  \left| a(l) \right| \right) \| f \|_{W^{k, 1}(\R^d)}.
\end{equation}
Next, recall that the Cesàro means, defined through
\begin{equation*}
    \sigma_N(x) = \sum_{\substack{l \in \Z^d \\[0.1cm] |l_j| \leq N}} \left( 1 - \frac{|l_1|}{N+1} \right) \cdots \left( 1 - \frac{|l_d|}{N+1} \right) \hat{p}(l) \exp(2 \pi i l^\top x), \quad x \in \T, \quad N \in \N,
\end{equation*}
are trigonometric polynomials whose coefficients are evidently bounded by the absolute value of the Fourier coefficients of $p$. So, using \eqref{eq:estimate_trig_poly}, we have the following uniform bound in $N$:
\begin{equation*}
    \left| \int_{\R^d} f(x) \left[ \exp \left( \sigma_N\left( x/\eps \right) \right) - \int_{\T} \exp \left( \sigma_N(y) \right) dy  \right] dx \right| \leq \frac{\eps^k }{(2\pi)^k} \exp \left( \sum_{l \in \Z^d}  \left| \hat{p}(l) \right| \right) \| f \|_{W^{k, 1}(\R^d)}.
\end{equation*}
Using this uniform bound and the triangle inequality, we eventually arrive at
\begin{align*}
    \left| \int_{\R^d} f(x) \left[ \exp \left( p\left( x/\eps \right) \right) - \int_{\T} \exp \left( p(y) \right) dy  \right] dx \right| 
    &\leq 2 e \| f \|_{L^1(\R^d)} \| p - \sigma_N \|_{C(\T; \R)} \\
    &\hspace{0.5cm} + \frac{\eps^k }{(2\pi)^k} \exp \left( \sum_{l \in \Z^d}  \left| \hat{p}(l) \right| \right) \| f \|_{W^{k, 1}(\R^d)}\,.
\end{align*}
It is a well-known fact, see \cite[Chapter 3]{G:2008}, that the Cesàro means converge to $p$ in $C(\T; \R)$ as $N \rightarrow \infty$ so that we finally arrive at the concluding estimate \eqref{eq:oscillatory_integral_estimate}. 
\end{proof}

\begin{Corollary}   \label{cor:gibbs_measure_estimates}
Let $k \in \N$. Consider a potential $V \colon \R^d \to \R$ as per Definition \ref{def:potential} and a $1$-periodic function $p \in C^2(\T; \R)$. Define the Lebesgue-densities
\begin{equation}
        \mu_\eps(x) = \frac{1}{Z_\eps} \exp\left(-V(x) - p\left(x/\eps\right)\right), \quad
        \mu_\eps(x) = \frac{1}{Z} \exp\left(-V(x)\right), \quad x \in \R^d, \quad \eps > 0,
\end{equation}
where $Z, Z_\eps > 0$ are normalization constants and set two random variables $\xi_\eps \sim \mu_\eps \, d \lambda^d$, $\xi \sim \mu \, d \lambda^d$. Then for any function $g \in C^k(\R^d; \R)$ that grows at most polynomially, it holds for sufficiently small $\eps > 0$
\begin{equation}
    \left| \E g(\xi_\eps) - \E g(\xi) \right| \leq C(k, p, V, g) \eps^k,
\end{equation}
with a constant $C(k, p, V, g) > 0$ which only depends on $k, p, V$ and $g$.
\end{Corollary}
\begin{proof}
Set 
\begin{equation}
    Z_p := \int_{\T} \exp(-p(y)) \, dy.
\end{equation}
Notice that
\begin{align*}
    \E g(\xi_\eps) - \E g(\xi) 
    &= \int_{\R^d} g(x) \left[ \mu_\eps(x) - \mu(x) \right] dx \\
    &= \int_{\R^d} g(x) \left[ \mu_\eps(x) - \frac{Z_p}{Z_\eps} \exp(-V(x)) \right] dx \\
    &\hspace{0.5cm} + \int_{\R^d} g(x) \left[ \frac{Z_p}{Z_\eps} \exp(-V(x)) - \mu(x) \right] dx \\
    &= \frac{1}{Z_\eps} \int_{\R^d} g(x) \exp(-V(x)) \left[ \exp(-p(x/\eps)) - Z_p \right] dx \\
    &\hspace{0.5cm} + \frac{Z Z_p - Z_\eps}{Z Z_\eps} \int_{\R^d} g(x) \exp(-V(x)) \, dx.
\end{align*}
Now observe that
\begin{equation}
    Z_\eps - Z Z_p = \int_{\R^d} \exp(-V(x)) \left[ \exp(-p(x/\eps) - \int_{\T} \exp(-p(y)) \, dy \right] dx.
\end{equation}
The Fourier series of $p$ is absolutely summable since $p \in C^2(\T; \R)$ and, thus, $|\hat{p}(l)| \sim l^{-2}$ for $l \in \Z^d$. Remark \ref{rem:V_potential_estimate} and Proposition \ref{prop:oscillatory_integral_estimate} tell us that
\begin{equation}
    |Z_\eps - Z Z_p| \leq \frac{\eps^k}{(2\pi)^k} \exp\left( \sum_{l \in \Z^d} |\hat{p}(l)| \right) \| \exp(-V) \|_{W^{k, 1}(\R^d)}.
\end{equation}
In particular, $Z_\eps \geq (Z Z_p)/2 > 0$ for sufficiently small $\eps > 0$.  Furthermore, using the estimate \eqref{eq:potential_exp_estimate} and the assumption that $g$ grows at most polynomially, it holds $x \mapsto g(x) \exp(-V(x)) \in S^k(\R^d)$. Therefore, Proposition \ref{prop:oscillatory_integral_estimate} eventually yields
\begin{align*}
    \left| \E g(\xi_\eps) - \E g(\xi) \right| 
    &\leq \frac{2 \eps^k}{Z Z_p (2\pi)^k} \exp\left( \sum_{l \in \Z^d} |\hat{p}(l)| \right) \| g \exp(-V) \|_{W^{k, 1}(\R^d)} \\
    &\hspace{0.5cm} + \frac{2 \eps^k}{Z^2 Z_p (2\pi)^k} \exp\left( \sum_{l \in \Z^d} |\hat{p}(l)| \right) \| \exp(-V) \|_{W^{k, 1}(\R^d)} \| g \exp(-V) \|_{L^1(\R^d)} \\
    &= C(k, p, V, g) \eps^k,
\end{align*}
for sufficiently small $\eps > 0$.
\end{proof}
The next result justifies the claim at the end of the proof of Proposition \ref{prop:asy_normal_langevin} that condition (2.78) of \cite[Theorem 2.19]{BK:2025} is satisfied. For this, define 
\begin{equation} \label{eq:poisson_langevin_eps}
    \Phi_\eps(x) := \int_0^x \frac{1}{\sigma \mu_\eps(\psi(\vt_0), y)} \int_{-\infty}^y h_\eps(z) \mu_\eps(\psi(\vt_0), z) \, dz \, dy, \quad x \in \R,
\end{equation}
with $\vt_0 \in \Theta_0$ and the invariant densities $\mu_\eps(\psi(\vt_0), \cdot)$ and $\mu(\vt_0, \cdot)$ defined as in Section \ref{sec:langevin}.
\begin{Lemma}   \label{lem:asymptotic_variance_convergence}
Let $\vt_0 \in \Theta_0$. In the multiscale overdamped Langevin drift parameter estimation problem, it holds
\begin{equation}    \label{eq:asymptotic_variance_convergence}
    \sigma \int_\R |\Phi_\eps'(x)|^2 \mu_\eps(\psi(\vt_0, x) \, dx \rightarrow \Bar{\sigma} \int_\R |\Phi'(x)|^2 \mu(\vt_0, x) \, dx, \quad \epstozero.
\end{equation}
\end{Lemma}
\begin{proof}
Using the Poisson equations
\begin{equation}    \label{eq:poisson_equations}
    \alpha V' \Phi_\eps' + \frac{1}{\eps} p'\left( \frac{\cdot}{\eps} \right) \Phi_\eps' - \sigma \Phi_\eps'' = h_\eps, \quad \alpha K V' \Phi' - \sigma K \Phi'' = h, \quad \text{on } \R,
\end{equation}
and integration by parts shows that
\begin{align}   \label{eq:dirichlet_form}
    \sigma \int_\R |\Phi_\eps'(x)|^2 \mu_\eps(\psi(\vt_0, x) \, dx &= \int_\R \Phi_\eps(x) h_\eps(x) \mu_\eps(x) \, dx, \\
    \overline{\sigma} \int_\R |\Phi_\eps'(x)|^2 \mu_\eps(\vt_0, x) \, dx &= \int_\R \Phi(x) h(x) \mu(x) \, dx.
\end{align}
Moreover, with Lemma \ref{lem:weak_conv_densities_langevin} and verbatim arguments that were used to prove \cite[Lemma 3.1, (3.8)]{BK:2025}, one can show for arbitrary $M > 0$ that
\begin{equation}    \label{eq:Phi_estimate}
    \sup_{x \in [-M, M]} |\Phi_\eps(x) - \Phi(x)| \rightarrow 0, \quad \text{as } \epstozero.
\end{equation}
Next, we choose a sufficiently large $R>0$ such that the integral
\begin{equation}    \label{eq:Phi_integral_tail}
    \int_{[-R, R]^c} |\Phi_\eps(x) - \Phi(x)| \,  \mu(x) \, dx 
\end{equation}
gets arbitrarily small uniformly in $\eps$. This is possible because the integrand is uniformly integrable by \cite[Lemma 2.18]{BK:2025}.
We can perform the following splitting
\begin{align*}
    \int_\R \Phi_\eps(x) h_\eps(x) \mu_\eps(x) \, dx - \int_\R \Phi(x) h(x) \mu(x) \, dx 
    &=  \int_\R \Phi_\eps(x) (h_\eps(x) - h(x)) \mu_\eps(x) \, dx \\
    &\hspace{0.35cm} + \int_\R (\Phi_\eps(x) - \Phi(x)) h(x) \mu_\eps(x) \, dx \\
    &\hspace{0.35cm} + \int_\R \Phi(x) h(x) (\mu_\eps(x) - \mu(x))\, dx.
\end{align*}
The first term vanishes by virtue of Lemma \ref{lem:weak_conv_densities_langevin} and Lemma \cite[Lemma 2.18]{BK:2025}. The third term vanishes due to weak convergence of $\mu_\eps \lambda^1$ and uniform integrability of $\Phi h$ with respect to $\mu_\eps \lambda^1$. The second term can be estimated as follows
\begin{align*}
    \left| \int_\R (\Phi_\eps(x) - \Phi(x)) h(x) \mu_\eps(x) \, dx \right| = 2 \left[ \sup_{x \in [-R, R]} |\Phi_\eps(x) - \Phi(x)| + \int_{[-R,R]^c} |\Phi_\eps(x) - \Phi(x)| \, \mu_\eps(x) \, dx \right].
\end{align*}
These terms go to zero by \eqref{eq:Phi_estimate} and \eqref{eq:Phi_integral_tail}, so that \eqref{eq:asymptotic_variance_convergence} is proved.
\end{proof}

\section*{Acknowledgements}
The authors J.I.B.\ and S.K.\ acknowledge funding by the Deutsche Forschungsgemeinschaft (DFG, German Research Foundation) - Project number 442047500 through the Collaborative Research Center "Sparsity and Singular Structures" (SFB 1481). G.P. is partially supported by an ERC-EPSRC Frontier Research Guarantee through Grant No. EP/X038645, ERC Advanced Grant No. 247031 and a Leverhulme Trust Senior Research Fellowship, SRF$\backslash$R1$\backslash$241055. The authors would like to express their gratitude to Mr. Maximilian Kruse for his support in setting up some of the computer experiments.

\bibliography{ref}

\end{document}